# Scalar Supersymmetry in Bi-Spinor Gauge Theory

Alex Jourjine
FGCTP
Hoffmann Str. 6-8
01281 Dresden
Germany

**Abstract**

We describe a new realization of both global and local supersymmetry acting in spaces of commuting and anticommuting differential forms. Unlike the standard supersymmetry, it has Lorentz scalar transformation parameters. It is related but is not reducible to standard supersymmetry. Reformulation of the Standard Model with the new supersymmetry, called scalar supersymmetry, can be achieved with the particle content of the SM. BRST symmetry is extended to include scalar supersymmetry multiplets. Linear realization of scalar supersymmetry with free fields or with fields interacting with background gravity is described. Gauge interactions require non-linear realizations of scalar supersymmetry and, except for *SU(2)*, non-linear gauge-fixing conditions. Requiring scalar supersymmetry of interacting action with the simplest chiral multiplet can reduce the dynamical content of the theory to that of the SM: one complex scalar, gauge fields, and three generations of Weyl spinors. At low energies scalar supersymmetry is explicitly broken by gauge interactions. However, in the asymptotically free case it becomes exact in the ultraviolet limit. Thus it has the two most desirable features of softly broken standard supersymmetry built in. Implementation of exact scalar supersymmetry in an interacting string action is given.

## 1. Introduction

It is well-known that the use of supersymmetry [1] to extend the Standard Model (SM) results in a number of attractive features. Apart from curing the Higgs mass finetuning problem, supersymmetry leads to gauge coupling unification, provides candidates for Dark Matter, and sets the stage for gravity unification via superstrings. Unbroken supersymmetry requires that each observed particle has a superpartner with equal mass. Since the observed particle mass spectrum of the SM is not mass-degenerate, supersymmetry must be broken. Breaking supersymmetry is a non-trivial problem, for it must be broken softly in order to preserve the ultraviolet properties of supersymmetry [2, 3, 4, 5].

One characteristic feature of all supersymmetric extensions of the SM appears in the particle spectrum even if supersymmetry is broken. Namely, each SM particle must have a superpartner with spin differing by one-half. This is because in the SM bosonic gauge fields are real and transform in the adjoint representation of the gauge group $G_{SM} = SU(3)_C \times SU(2)_L \times U(1)_Y$ but fermionic spinor fields are complex and transform in the fundamental representations of $G_{SM}$. As a result, one cannot combine the observed bosons and fermions into multiplets without violating gauge symmetry. In addition, left- and right-handed fermions couple differently to $SU(2)_L$. To accommodate the difference one is forced to use chiral supermultiplets. These can be only constructed if one pads each fermion or boson of the SM with a superpartner particle of differing spin.

Despite an intensive search at LHC, no superpartner particles of the Standard Model particle spectrum have been detected [6, 7, 8]. Of course, one can explain the failure to observe them by making the superpartner particles sufficiently heavy and by adjusting a fairly large number of parameters that even the simplest realistic supersymmetric extensions of the SM bring with them. However, another explanation for the failure could be that the superpartners don't exist at least for some part of the spectrum of the SM. An explanation of this sort would require a different realization of supersymmetry.

In this Letter we describe such a realization of supersymmetry. It does not require doubling of the particle spectrum, because the observed bosonic and fermionic fields are allowed to form chiral supersymmetric multiplets. In fact, the particle content of the SM reformulated in terms of supersymmetry realized with bi-spinors can remain as it is. Just like in the standard supersymmetry the additional field content of such supersymmetrization is completely determined by supersymmetry requirement. In this regard the SM reformulated in terms of the new supersymmetry is no more arbitrary then the standard supersymmetric models currently under investigation.

Because the anticommuting transformation parameters of the new realization are Lorentz scalars, we call it scalar supersymmetry (s-supersymmetry). We show that s-supersymmetry can always be realized linearly as symmetry of free action or non-linearly as symmetry of interacting action. Non-linear realizations with $SU(1)$ and $SU(3)$ require non-linear Lorentz gauge-fixing condition. Non-linear realization with $SU(2)$ can be done with linear Lorentz gauge-fixing condition.

We show that in its linear realization s-supersymmetry is a partial symmetry. That is it is symmetry of only a part of the full quantum BRST-symmetric action; gauge interactions break it. We consider this to be an advantage because s-supersymmetry is restored in the ultraviolet limit in asymptotically free theories. In any case, partial symmetries are nothing new in particle physics. Breakdown of s-supersymmetry due to interactions is similar to breakdown of chiral symmetry by interaction induced by mass terms.

S-supersymmetry can be realized on an arbitrary space-time manifold and in the presence of background gravity s-supersymmetry involves metric in a non-trivial way. It this aspect it fundamentally differs from the standard supersymmetry, because it can be defined on any smooth space-time manifold, even when spinors and hence the standard supersymmetry cannot be defined consistently.

Unlike the standard supersymmetry, s-supersymmetry acts in the space that is a direct sum of spaces of commuting and anti-commuting differential forms (difforms) with values in a Lie algebra representation. It requires the use of bi-spinor[1] formalism to represent fermionic matter. As a result, in bi-spinor formalism bosonic and fermionic fields are described by the same objects, that is, difforms on a manifold.

---

[1] Bi-spinors are also referred to as Ivanenko-Landau-Kähler spinors [56], which is historically most appropriate, but most often as Dirac-Kähler spinors.

The only difference between them is that bosonic field difforms commute, while fermionic field difforms anticommute. Thus, in bi-spinor formalism the difference between fields of different spin is reduced to absolutely essential minimum.

The unlike appearance of bosonic fields and spinors in some forms of bi-spinor actions can be traced to the fact that, mathematically, bi-spinors are coefficients of expansion of fermionic difforms in a certain spinor difform basis, called the *Z*-basis, while bosonic fields are coefficients of expansion of difforms in the usual coordinate basis, the *c*-basis. However, bosonic fields can as well be converted into the *Z*-basis, just like fermions can be converted into the *c*-basis. We will present examples of where such conversion makes expressions more tractable.

Algebraically bi-spinors are sums of products of Dirac spinors and their Dirac conjugates, but only one half of Dirac spinor factors of bi-spinors carry the physical degrees of freedom of bi-spinors. Each physical factor can be identified as a single generation of fermion particles. As we will show in Appendix A the number of generations can be made arbitrary.

The notion of bi-spinor and its use in Physics is as old as that of Dirac spinor. In its anti-symmetric tensor form (in the *c*-basis) it was first discovered and used by Ivanenko and Landau in 1928 [9], in the same year Dirac proposed his revolutionary theory of electron [10]. Using bi-spinors Ivanenko and Landau constructed an alternative to Dirac's solution of the electron's gyromagnetic ratio problem. Naturally, Dirac's solution won over, since it was much simpler.

Only much later in 1962 the appropriate mathematical setting for bi-spinors in terms of differential forms on space-time manifolds was established in [11] and further elucidated in [12, 13]. Bi-spinors have not been popular in phenomenological model building but they have been much in use in lattice gauge theory and, in particular, for building realizations of Dirac-Kähler twisting of the standard extended supersymmetry on the lattice [14, 15, 16, 17, 18]. We will discuss the connection between s-supersymmetry and twisted Dirac-Kähler supersymmetry below. Antisymmetric tensor form of bi-spinors appears quite often in string theories in the form of c-basis differential forms of fixed degree, the *p*-forms. *P*-forms and their quantization have been studied both in supergravity and in string theory, including formulation of strings with two time parameters, where such objects appear naturally [19, 20, 21, 22, 23]. However, theories of *p*-forms typically are restricted to commuting differential forms of a fixed degree. Here we will concentrate on applications to Beyond the SM phenomenology, where it is the inhomogeneous differential forms that play the central role.

There were good reasons why bi-spinors were not used in phenomenologically realistic models beyond the SM even after it was realized in [24, 25] that they can describe multiple generations of lepto-quarks. First, for a long time one could apply the standard Dirac field quantization rules only on Euclidean space-times, which is of course sufficient only for lattice applications. On pseudo-Euclidean space times, such as the observed Minkowski space-time, one cannot quantize bi-spinors using the standard Dirac quantization procedure without generating unphysical modes [26, 27]. This problem has been resolved only recently by modifying the Dirac quantization rules for some of the modes that in essence exchange the definition of particle and anti-particle for the modes [28].

Second, it is commonly assumed that bi-spinor formalism requires exactly four generations of lepto-quarks. This turns out to be not a problem. We will show below in the Appendix A how bi-spinors can describe arbitrary number of generations.

Third, in the presence of gravity bi-spinors describe physical objects that are distinctly different from the familiar Dirac spinors. It is well-known that Einstein-Hilbert gravity cannot incorporate matter with half-integer spin. The use of Dirac spinors in general relativity requires the Cartan formulation of gravity, where the primary variables are frames and spin connection, while metric is a derived quantity. Unlike Dirac spinors that must couple to spin connection of the Cartan formulation of gravity, bi-spinors can be incorporated into Einstein-Hilbert gravity, for they couple directly to metric.

One of the first attempts to rewrite the SM in terms of bi-spinors was carried out in [29], where quaternion valued antisymmetric tensor fields were used as the basic variables for gauge fields. In our formulation the use of quaternions is not necessary. In addition, in [29] a different form of bi-spinor conjugation was used to construct invariant actions. The choice assured that free action can be reduced to the standard Dirac action for four generations of Dirac spinors. However, while acceptable on Minkowski or Euclidean space-times, the conjugation is not generally covariant, because it introduces a specific time direction. Another, attempt, this time with the generally covariant form of conjugation, was made in [30]. However, there the mass terms that were used in the action were not of the most general form. As a result, one obtains a degenerate mass spectrum.

Recently there appeared indications that a reformulation of the SM based on bi-spinors that uses most general possible mass terms contains an additional quantum number for elementary particles, the scalar spin [31, 32]. Scalar spin appears when bi-spinor degrees of freedom are reduced to Dirac spinor degrees of freedom. It appears because the two matrices arising from the bilinear terms for the kinetic and the mass terms of the free bi-spinor action, in general, do not commute. As a result, the usual mass matrix diagonalization procedure of the SM does not work and admissible mass terms become severely restricted. One consequence of the presence of scalar spin in a bi-spinor gauge theory is essentially unique textures of lepto-quark mixing matrices $V_{CKM}$ and $U_{PMNS}$. Details of the quantization of bi-spinors in bi-spinor gauge theory and detailed discussion of possible mass terms and of scalar spin can be found in [33].

As a somewhat tangential topic, we also present a realization of the exact scalar supersymmetry in an example of interacting string action. We have included this material to prove the point that scalar supersymmetry does not have to be broken by interactions. An additional motivation for considering s-supersymmetric string is that there always exists a possibility that a new formulation of superstrings can bring some advantages and new insights to the standard string theory. Whether there exist a linear realization of unbroken scalar supersymmetry for interacting bi-spinor gauge theory is not clear. Our point of view is that, as far as building realistic models is concerned, unbroken supersymmetry in not needed, the notion being unsupported by the experiment. It is the broken supersymmetry that is needed for models beyond the SM.

To conclude the introduction we would like to emphasize that the purpose of this Letter, which is built on the results of free field supersymmetry discussed in [34], is not to construct a new supersymmetric reformulation of the Standard Model in terms of bi-spinors. The Letter is only an explorative work to find out the advantages and the disadvantages of the use of scalar supersymmetry for constructing phenomenological models. We leave a detailed formulation of the bi-spinor SM to a future publication.

The Letter has six sections and two Appendixes. In Section 2 we rewrite the conventional gauge theory in coordinate- and Lie algebra basis-free fashion in terms of standard operations on difforms, describe how to rewrite the standard BRST symmetry for the bi-spinor case, and show how to extend it to the s-supersymmetric case. In Section 3 we describe the simplest implementations of s-supersymmetry for the $U(1)$ gauge group case. It contains examples of s-supersymmetry of free action with a complex and chiral multiplets and description of the consequences of non-linear s-supersymmetry of the interacting action. This is followed in Section 4 with extension of the results to $SU(N) \times U(1)$ and, with an eye on the SM, to the $SU(N_3) \times SU(N_2) \times U(1)$ cases. Section 5 very briefly discusses s-supersymmetric string. It is somewhat tangential section but is important as an example of implementation of exact s-supersymmetry in an interacting theory. Section 6 is a summary. In Appendix A we describe the conventions, the needed basic ingredients of differential geometry, extraction of algebraic (anti-)Dirac[2] spinors from bi-spinors, and the covariant constraints needed to set the number of generations that the bi-spinors contain to the desired number. Tables of commutation properties of basic operators in bi-spinor gauge theory are contained in Appendix B.

## 2. Differential Geometry, Gauge Theory, and Bi-Spinor BRST

In this section we will describe the standard and the bi-spinor-gauge theories with their BRST transformations in terms of basic notions of differential geometry. The standard description of $U(1)$ or $SU(N)$ gauge fields in quantum field theory is by a $g$–valued connection $A_\mu$, $A_\mu = A_\mu^a \tau^a$, $A_\mu \in g$, where $\tau^a = \tau^{ab}{}_F$ are $M$ matrices of the fundamental representation of $g$, with normalization described in (A.40-44). Large gauge transformations of the connection and its curvature $F_{\mu\nu}$ are defined by

$$A_\mu \to \Omega\, A_\mu \Omega^{-1} + \frac{i}{g} \Omega\, \partial_\mu \Omega^{-1}, \qquad F_{\mu\nu} \to \Omega\, F_{\mu\nu} \Omega^{-1}, \qquad \Omega = \exp\!\left(i\omega^a \tau^a{}_F\right). \quad (2.1)$$

Fermionic degrees of freedom are usually described by a collection of $N_g$ generations of Dirac spinors $\psi^{aA}$, $A = 1,\ldots,N_g$, transforming in possibly different representations of $g$ for left- and right-handed fermions, but representation assignment is assumed to be generation-independent. Since, as we will see below, scalar supersymmetry requires that $\psi$ transforms in the fundamental representation of $g$ as

$$\psi^A \to \Omega\, \psi^A, \qquad (2.2)$$

we will restrict ourselves to this particular case and drop index $F$ in $\tau^{ab}{}_F$.

In the $\xi$-gauge the quantum gauge-fixed Lagrangian with Faddeev-Popov ghost terms are given by

$$S = \int d^4x\, \mathcal{L},$$

(2.3)

---

[2] The meaning of the term anti-Dirac spinor is described at the end of Section 2.

$$\mathcal{L} = -\frac{1}{4}\left(F_{\mu\nu}{}^a F^{\mu\nu a}\right) + \left(\overline{\psi}^{iA}\left(i\gamma^\mu D_\mu^{ik}\right)\psi^{kA}\right) - \frac{1}{2\xi}\left(\partial_\mu A^{\mu a}\partial_\mu A^{\mu a}\right) - \left(\overline{c}^{\,a}\partial^\mu D_\mu^{ab} c^b\right),$$

where $F_{\mu\nu} = F_{\mu\nu}^a \tau^a$, the curvature of the connection $A_\mu$, is related to gauge field $A_\mu$ by

$$F_{\mu\nu}^a = \partial_\mu A_\nu^a - \partial_\nu A_\mu^a + g f^{abc} A_\mu^b A_\nu^c, \tag{2.4}$$

where $g$ is the coupling constant, and the Lagrange $\xi^{-1}$ multiplier fixes the gauge. Ghost fields $\overline{c}^{\,a}$ and $c^a$ are two unrelated real anticommuting fields. Note the difference in dimension of spaces where the two covariant derivatives in (2.3) act in. When acting on spinors the covariant derivative $D_\mu^{ik}$ acts in $N$-dimensional complex space and is given by

$$D_\mu^{ik} = \partial_\mu \delta^{ik} - ig A_\mu^a \left(\tau_F^a\right)^{ik}, \quad i,k = 1,\ldots,N. \tag{2.5}$$

When acting on gauge fields and ghosts the covariant derivative $D_\mu^{ab}$ acts in $(N^2-1)$-dimensional space of traceless hermitean complex $N\times N$ matrices, that is in the $(N^2-1)$-dimensional adjoint representation

$$D_\mu^{ab} = \partial_\mu \delta^{ab} + g f^{acb} A_\mu^c, \quad a,b,c = 1,\ldots,(N^2-1), \tag{2.6}$$

whereas for the $U(1)$ case we have

$$D_\mu = \partial_\mu - ig A_\mu. \tag{2.7}$$

We will see below that (2.5) and (2.6) become identical in bi-spinor formalism. The first two terms of Lagrangian (2.3) are invariant under gauge transformations (2.1-2) or, infinitesimally, for $\Omega = 1 + ig\varepsilon\omega^a\tau^a + O(\varepsilon^2)$

$$\delta_\varepsilon \psi^{aA} = ig\varepsilon\omega^c \left(\tau^c\right)^{ab} \psi^{bA}, \tag{2.8}$$

$$\delta_\varepsilon A_\mu^a = \varepsilon\left(\partial_\mu \omega^a + g f^{abc} A_\mu^b \omega^c\right), \tag{2.9}$$

$$\delta_\varepsilon F_{\mu\nu}^a = -g\varepsilon f^{abc} \omega^b F_{\mu\nu}^c. \tag{2.10}$$

Although Lagrangian (2.3) by construction is not gauge invariant, using additional transformations of ghost fields, the partial gauge invariance of (2.3) under (2.1-2) can be enlarged to invariance under the BRST symmetry transformations given by

$$\delta_\varepsilon \psi^{aA} = ig\varepsilon c^c \left(\tau^c\right)^{ab} \psi^{bA},$$

$$\delta_\varepsilon A_\mu^{\ a} = \varepsilon\left(\partial_\mu c^a + g f^{abc} A_\mu^b c^c\right) = \varepsilon D_\mu^{ac} c^c,$$

$$\delta_\varepsilon c^a = -\frac{1}{2} g \varepsilon f^{abc} c^b c^c,$$

$$\delta_\varepsilon \bar{c}^a = -\frac{1}{\xi}\varepsilon \partial^\mu A_\mu^a,$$

(2.11)

where now $\varepsilon$ is an infinitesimal anticommuting parameter[3] and, as before, $\bar{c}^a$ and $c^a$ are unrelated real anticommuting (Grassmann) fields. The first two transformations in (2.11) are, in fact, the usual gauge transformations (2.8-9) of the gauge invariant part of the Lagrangian with commuting but nilpotent parameters $(\varepsilon c^a)$. The most important property of BRST transformation (2.11) is its nilpotency for arbitrary $\xi$: $\delta_\varepsilon^2 \equiv 0$. This property is essential for proving unitarity of $S$-matrix in arbitrary $\xi$-gauge.

We will now rewrite the Lagrangian and the BRST transformations in terms of coordinate- and Lie algebra basis-invariant operators acting on difforms, thus removing from their description all references to a particular basis in $\mathfrak{g}$ and to Dirac spinors. First we will introduce the standard commuting connection 1-form $A_1$ by $A_1 = A_\mu dx^\mu$ and define the curvature 2-form $F$ by

$$F = (1/2) F_{\mu\nu} dx^\mu \wedge dx^\nu.$$

From (2.7) we obtain that

$$F = (d - i g A_1 \wedge) A_1 = d_{A_1} A_1, \qquad d_{A_1} \equiv (d - i g A_1 \wedge), \qquad (2.12)$$

where we used the gauged exterior derivative $d_{A_1}$ constructed from the exterior derivative $d$ defined in (A.4) and the exterior product on difforms $\wedge$. Operator $d_{A_1}$ is a non-Abelian generalization of the well-known in the electromagnetism relation $F = d A_1$, $F_{\mu\nu} = \partial_\mu A_\nu - \partial_\nu A_\mu$, where coefficients $A_\mu$ of $A_1$ describe the electromagnetic vector potential. From (A.11) we obtain that

$$\partial_\mu A^\mu = -\delta A_1. \qquad (2.13)$$

Consequently, using (A.40-44) we can write down Lagrangian (2.3) as

$$\mathcal{L} = tr(\alpha d_{A_1} A_1, d_{A_1} A_1) + \left(\bar{\psi}^{iA}\left(i\gamma^\mu D_\mu^{ik}\right)\psi^{kA}\right) - \frac{1}{\xi} tr(\alpha \delta A_1, \delta A_1) + 2 tr(\alpha \bar{c}, \delta d_{A_1} c), \qquad (2.14)$$

---

[3] For simplicity we use the original form of BRST. Our results extend as well to the conventional form, where additional non-dynamical $\mathfrak{g}$-valued field is used to include $\bar{c}^a$ in $\delta_\varepsilon^2 \equiv 0$ identity.

where we defined $\mathfrak{g}$–valued 0-forms $c = c^a \tau^a$, $\bar{c} = \bar{c}^a \tau^a$, used definition (A.2) of the main automorphism $\alpha$, and definition (A.3) of contraction of difforms. Note that, because large gauge transformation parameters $\Omega$ are 0-forms and, hence, $\Omega \wedge A \equiv \Omega \cdot A$, large gauge transformations (2.1) can be also written as

$$A_1 \to \Omega \wedge A_1 \wedge \Omega^{-1} + \frac{i}{g}\Omega \wedge d\,\Omega^{-1}, \quad F \to \Omega \wedge F \wedge \Omega^{-1}. \tag{2.15}$$

In terms of difforms $A_1$, $c$, $\bar{c}$ the BRST transformations (2.11) now take the form

$$\delta_\varepsilon \psi^A = ig\,\varepsilon\,c\psi^A,$$

$$\delta_\varepsilon A_1 = \varepsilon d_{A_1} c, \qquad d_{A_1} c = (d\,c - ig[A_1, c])$$

$$\delta_\varepsilon c = i g\,\varepsilon c^2, \tag{2.16}$$

$$\delta_\varepsilon \bar{c} = \frac{1}{\zeta}\varepsilon\,\delta A_1,$$

The only terms in (2.14, 2.16) that spoil our attempt to write the Lagrangian in terms of operators acting on difforms are the terms containing Dirac spinors. We can complete our program if instead of Dirac spinors we use bi-spinors as fundamental descriptors of fermionic matter[4]. To do this we use relation (A.32) between a generation multiplet of Dirac spinors, spinbeins, and bi-spinors and instead of $\psi^{iA}$ use $\Psi^{ik} = \psi^{iA}\bar{\bar{\eta}}^{Ak}$, where $\Psi^{ik}$ is a set of coefficients of the inhomogeneous differential form $\Phi^{ik}$ in the $Z$-basis, i.e., $\Phi^{ik} = tr(Z\,\Psi^{ik})$. Since spinbein $\bar{\bar{\eta}}^{Ak}$ is a commuting entity and Dirac spinors $\psi^{iA}$ are anticommuting, the coefficients of difform $\Phi^{ik}$ must be taken to be anticommuting. Accordingly we now replace the Dirac spinor part of Lagrangian (2.14) following successive substitutions given by

$$\left(\bar{\psi}^{iA}\left(i\gamma^\mu D_\mu^{ik}\right)\psi^{kA}\right) \to \left(\bar{\bar{\psi}}^{iA}\left(i\mathbb{D}^{ik}\right)\psi^{kA}\right) = tr\left(\bar{\bar{\Psi}}(i\mathbb{D})\Psi\right), \tag{2.17}$$

where we used (A.30-32). Note that during transition from $\psi^{kA}$ to $\Phi^{ik}$ we enlarged the dimension of the space where the covariant derivatives $D_\mu^{ik}$ act to $2N^2$, the space which contains the $(N^2 - 1)$- dimensional space of the adjoint representation and 1-dimensional space of the trivial representation. Thus $D_\mu^{ik}$ in (2.5) acting on $\Phi^{ik}$ restricted to adjoint representation space is identical to $D_\mu^{ab}$ in (2.6). The apparent difference between the two is the difference in choosing the Lie algebra basis for the adjoint representation.

Using (2.17, A.11) we obtain the fermionic part of the Lagrangian in (2.14) in the desired differential-geometric form in terms of the anti-commuting difform $\Phi$

$$\mathcal{L}_f = tr\left(\alpha\,\Phi, (d_{A_1} - \delta_{A_1})\Phi\right), \quad \delta_{A_1} = *\,d_{A_1}\,*, \tag{2.18}$$

---

[4] To distinguish them from the standard gauge theories, we call such theories bi-spinor gauge theories.

where $-\delta_{A_1}$ is the adjoint of $d_{A_1}$ with respect to scalar product (A.14). The explicit form of $\delta_{A_1}$ acting on $A_1$ will play an important role below. It is given by

$$\delta_{A_1} A_1 = -\partial^\mu A_\mu + ig\, A^\mu A_\mu \qquad \text{for} \quad U(1),$$

$$\left(\delta_{A_1} A_1\right)^a = -\partial^\mu A_\mu^{\;a} \qquad \text{for} \quad SU(2), \qquad (2.19)$$

$$\left(\delta_{A_1} A_1\right)^a = -\partial^\mu A_\mu^{\;a} + i\frac{g}{4} d^{abc} A_\mu^{\;b} A^{\mu c} \qquad \text{for} \quad SU(N),\ N\geq 3.$$

As for BRST, we see that, except for the first one, all BRST transformations in (2.11) are already in the desired invariant form. Since BRST does not mix generations we can use the linearity of spinbein decomposition (A.32) and of difform decomposition (A.18) to make replacements

$$ig\,\varepsilon c\psi^{iA} \to ig\,\varepsilon c\Psi^{ik}(\Phi) \to ig\,\varepsilon c\Phi^{ik}. \qquad (2.20)$$

We thus arrive at the $\xi$-gauge quantum gauge-fixed Lagrangian and its BRST symmetry transformations expressed entirely in terms of differential-geometric operators acting on difforms

$$\mathcal{L}=tr\left(\alpha d_{A_1} A_1, d_{A_1} A_1\right)+tr\left(\alpha\Phi,(d_{A_1}-\delta_{A_1})\Phi\right)-\frac{1}{\xi}tr(\alpha\delta A_1,\delta A_1)+2tr(\alpha\bar{c},\delta d_{A_1} c),$$

$$\delta_\varepsilon \Phi = ig\,\varepsilon c\Phi, \qquad \delta_\varepsilon A_1 = \varepsilon d_{A_1} c, \qquad (2.21)$$

$$\delta_\varepsilon c = ig\,\varepsilon c^2, \qquad \delta_\varepsilon \bar{c} = \frac{1}{\xi}\varepsilon \delta A_1.$$

This Lagrangian, as was sought, contains no reference to a specific Lie algebra basis or to Dirac spinors. Note that by replacement $\psi^{iA} \to \Phi^{ik}$ we went from a set of Dirac spinors $\psi^{iA}$, $A=1,\ldots,N_g$, each transforming in the fundamental representation of $SU(N)$ to a single bi-spinor $\Phi^{ik}$ transforming in the $N\times\bar{N}$ of $SU(N)$. $\Phi^{ik}$ form a linear space of complex $4N\times 4N$ matrices of rank $N_g\times N$. Important to note is that for arbitrary spinbeins fermionic Lagrangian (2.18) is invariant when $\Phi$ transforms under global $N\times\bar{N}$ of $SU(N)$. However, for local gauge transformations with a choice of a particular physical spinbein gauge with constant spinbein, the invariance reduces to the invariance corresponding to the fundamental representation of $SU(N)$.

Note also that, written as $\delta_\varepsilon c = (ig\,\varepsilon c)\cdot c$, the transformation for ghost $c$ in (2.21) is the same in its form as for bi-spinor $\Phi$. One advantage of the covariant form of BRST in (2.21) is that one does not have to deal with expressions involving structural constants, which simplifies some calculations. For example, proving the BRST invariance of the $d_{A_1} c$ term in (2.21) now relies on two obvious equations

$$dc^2 = dc \cdot c + c \cdot dc,$$

$$[A_1, c^2] = [A_1, c] \cdot c + c \cdot [A_1, c].$$

It is easy to verify that replacement $ig\, \varepsilon\, c\psi^A \to ig\, \varepsilon\, c\Phi$ preserves the nilpotency of $\delta_\varepsilon \Phi$. Therefore, the bi-spinor BRST transformation (2.21) is nilpotent. When applied to any field we obtain respectively for the original BRST

$$\delta_\varepsilon^2 \Phi = \delta_\varepsilon^2 A_1 = \delta_\varepsilon^2 c = 0,$$

$$\delta_\varepsilon^2 \Phi = \delta_\varepsilon^2 A_1 = \delta_\varepsilon^2 c = 0.$$

(2.22)

The standard BRST is augmented with $\delta_\varepsilon^2 \bar{c}_1 = 0$.

Large gauge transformations (2.1) can also be cast into differential-geometric form in the Z-basis. Using the fact that for 0-forms $\Omega_0$ in (A.18) we have
$\Psi(\Omega_0) = \Omega_0$, $\Psi(\Omega_0 \cdot \Phi) = \Psi(\Omega_0) \cdot \Psi(\Phi)$, $\Psi(\Omega_0 \cdot A_1 \cdot \Omega^{-1}) = \Psi(\Omega_0) \cdot \Psi(\Phi) \cdot \Psi(\Omega^{-1})$,
$\Psi(\delta\Omega_0) = \Psi(0) = 0$, we obtain the Z-basis gauge transformations

$$\Psi(\Phi) \to \Psi(\Omega_0) \cdot \Psi(\Phi),$$

$$\Psi(A_1) \to \Psi(\Omega_0) \cdot \Psi(A_1) \cdot \Psi(\Omega_0)^{-1} + \frac{i}{g} \Psi(\Omega_0) \cdot (i\partial) \Psi(\Omega_0)^{-1},$$

$$F \to \Psi(\Omega_0) \cdot \Psi(F) \cdot \Psi(\Omega_0)^{-1},$$

where all products are matrix multiplications of matrices with representation and spinor indices. To emphasize that the parameter of gauge transformation is a 0-form we replaced $\Omega$ in (2.1) with $\Omega_0$.

Equations of motion derived from the gauge invariant part of the Lagrangian in (2.21) are the familiar inhomogeneous Maxwell equations. These are supplemented by algebraic identities, the well-known Bianchi identities, which, in fact, are the homogeneous Maxwell's equations. Altogether we obtain coordinate and basis-free form of (non)-Abelian Maxwell's equations

$$\delta_{A_1} d_{A_1} A_1 = J_1, \qquad J_1 = \sum_{p=0}^{3} (\alpha \Phi_{p+1}, \Phi_p), \qquad (d_{A_1} - \delta_{A_1}) \Phi = 0,$$

$$(d_{A_1})^2 A_1 = 0, \qquad (\delta_{A_1})^2 A_1 = 0, \qquad \delta_{A_1} J_1 = 0,$$

(2.23)

where the 1-form current $J_1$ is defined via difform contraction (A.3). The last equation in (2.23) is the covariant conservation of the current. It provides the integrability condition for the first equation. Its conservation follows from equations of motion and the Bianchi identities in (2.23). In the Z-basis and in the spinbein basis for $\Psi(\Phi)$ the equation of motion for difform $\Phi$ in (2.23) acquire a progressively familiar form

$$(i\partial + g\,\Psi(A_1))\Psi(\Phi) = 0, \qquad \Psi(A_1) = \gamma^\mu A_\mu,$$

$$\gamma^\mu(i\partial_\mu + g\,A_\mu)\psi^A(\Psi) = 0, \qquad A = 1,\ldots,N_g, \tag{2.24}$$

where $\Psi(\Phi)$ is the bi-spinor corresponding to difform $\Phi$, while $\psi^A(\Psi)$ is the (anti)-Dirac spinor multiplet of $N_g$ generations according to (A.18, 30-33) and the number of generations $N_g$ is defined as the rank of spinbein matrix, $N_g = rank\,\eta_\alpha^A$.

We will now make a short diversion to gravitation with the purpose of pointing out the physical difference between Dirac spinors and bi-spinors on curved space-times. For gravity with spin connection 1-form $\Omega_1$, $\Omega_1 = \Omega_\mu dx^\mu$, coframe 1-forms $e^I$, $e^I = e^I_\mu dx^\mu$, $I = 1,\ldots,4$, and tangent space $\gamma$–matrices $\gamma^I$, $\{\gamma^I,\gamma^J\} = 2\eta^{IJ}$, $\eta^{IJ} = diag(1,-1,-1,-1)$, for massless fermions instead of (2.24) one obtains [24, 25, 35]

$$(i\partial\Psi(\Phi) + g_G[\Omega,\Psi(\Phi)]) = 0. \tag{2.25}$$

To emphasize the connection between gravity and a bi-spinor gauge theory we introduced a dimensionless gravitational coupling constant $g_G$, which is not to be confused with the dimensionful Newton's constant $G_N$. It can be computed within the context of Cartan gravity theory where the curvature 2-form is the sum of the Einstein-Hilbert 2-form and curvature 2-form of anti-de Sitter space-time [36]. One obtains

$$g_G = \left(\frac{8}{3}\pi\frac{\hbar\,G_N\Lambda}{c^3}\right)^{\frac{1}{2}} = \left(\frac{8}{3}\pi\Lambda\right)^{\frac{1}{2}} L_{Pl} \approx 0.5\cdot 10^{-60}, \tag{2.26}$$

where $L_{Pl} = (\hbar\,G_N/c^3)^{1/2}$ is the Plank length.

The use of spinbein decomposition results in a new form of this equation describing the coupling to gravity of (anti)-Dirac spinors $\psi^{iA}$

$$\gamma^a e_a^\mu(i\partial_\mu + g_G\Omega_\mu)\psi^{iA} - g_G\tilde{\Omega}^{AB}\psi^{iB} = 0, \qquad \tilde{\Omega}^{AB} = \overline{\overline{\eta}}^B\Omega\eta^A. \tag{2.27}$$

This should be compared with the equation of motion for Dirac spinors $\psi^{iA}$ in Palatini formulation of gravity, where coframe (cotetrad) and spin connection are independent variables

$$\gamma^a e_a^\mu(i\partial_\mu + g_G\Omega_\mu)\psi^{iA} = 0. \tag{2.28}$$

For simplicity we assumed that in (2.25) the left- and the right-handed fermions have the same spinbein decomposition. Obviously, in a gravitational field the two objects are physically different. We see that in (2.27) spin connection 1-form $\Omega_a$ generates a mass-like term that mixes different generations of (anti-)Dirac spinors $\psi^{iA}$. On space-times with constant curvature the term becomes constant and can be considered as a mass term. We conclude that gravity can induce a small cosmological mass in (anti-)Dirac spinor components of bi-spinors. We refer for further details to [33].

The physical reason for difference of (2.24) and (2.25, 2.27) is that, as can be seen from (A.21), $\Psi(\Phi)$ transforms in the adjoint spinor representation of the group $SL(2,C)$ of local frame rotations as $\Psi(\Phi) \to S(\Lambda)\Psi(\Phi)S^{-1}(\Lambda)$, whereas it transforms in the fundamental representation of the gauge group $G$. As a result, the gravitational covariant derivative in (2.25) involves a commutator.

Returning to bi-spinor gauge theory on $M_4$, we see that the first two transformations in (2.21) are in fact gauge transformations of fermionic difform and the gauge field 1-form. Note that, because it contains a conjugated spinbein transforming in anti-fundamental representation, the difform $\Phi^{ik}$ is a $N \times N$ matrix in representation indices. Nevertheless, as already mentioned above, $\Phi^{ik}$ still transforms in the fundamental representation under local $SU(N)$ gauge transformations. The physical reason for this is that the choice of physical constant spinbein gauge fixes the vacuum in quantum theory and hence the Fock space. Once a constant spinbein is chosen one can apply to it only global gauge transformations, which leave it constant. This fact underscores the difference between quantum gravity on $M$ and quantum field theory on $M_4$ that possibly lies at the core of difficulties with quantization of gravity: The structure of Fock space of quantum field theory is not compatible with the general equivalence principle, the basis for construction of classical gravity. A clear example of the incompatibility is the well-known Unruh effect, where transition into a constantly accelerating frame replaces vacuum and one-particle states by thermal distributions with temperature proportional to acceleration [37, 38, 39, 40].

Comparing the roles that the gauge and the fermionic fields play in Lagrangian (2.21) we see that, apart of the order of the derivatives in the corresponding part of the Lagrangian, the principle and the only difference between bosonic fields and bi-spinor fermionic fields is their commutativity property. Otherwise, both are described by the same mathematical object, a differential form on a manifold transforming in some representation of the gauge group. Notably, on arbitrary manifold $M$ Dirac spinors, i.e., sections of Clifford bundle on $M$ [12], as mathematical objects are quite different from gauge fields. The distinction appears most clearly for space-times where Dirac spinors cannot be defined but which are physically perfectly acceptable for definition of gauge fields or fermionic differential forms. The problem with existence of Dirac fields is that their existence requires existence of spin structure, which in turn can only exist when certain characteristic classes of difforms on the space-time manifold vanish [41]. For the same reason differential forms with values in sections of Clifford bundle cannot be defined on arbitrary $M$. In contradistinction, both bosonic and fermionic difforms can be defined on any smooth $M$.

It should be pointed out that even on curved space-time manifolds where Dirac spinors can be defined they give rise to certain problems, such as problems with the derivation of hermitean Hamiltonian [42], with non-existence of classical point-like and string-like objects [43], and non-uniqueness of minimal coupling of fermions to Cartan gravity [44]. Most important for applications to cosmology, Dirac spinors cannot be used to describe fermionic matter in Einstein-Hilbert gravity, the formulation of gravity that is most often used in cosmological models.

We should emphasize that replacement in (2.17) of $N_g$ generations of Dirac spinors in (2.14) with a single difform $\Phi^{ik} = tr(Z\Psi^{ik})$ entering (2.21) may have non-trivial physical consequences for some scattering amplitudes. Consider the free part of fermionic Lagrangian with $N_g = 4$ generations in (2.21). It is given by

$$\mathcal{L}_f^0 = tr\, \overline{\overline{\Psi}}(i\partial)\Psi = tr\, \overline{\overline{\psi}}^A (i\partial)\psi^A, \qquad (2.29)$$

where $\overline{\overline{\Psi}} = \gamma^0 \Psi^+ \gamma^0$, $\overline{\overline{\psi}}^A = \Gamma^{AB}\overline{\psi}^A$, $\Gamma^{AB} = diag(+1,+1,-1,-1)$, are the bi-spinor conjugations of $\Psi$ and $\psi$, respectively. Lagrangian (2.29) is an alternating sum of Lagrangians for four Dirac spinors $\psi^A$, two of which, those with $A = 1, 2$, enter the sum with the plus sign, while spinors with $A = 3, 4$ enter with the minus sign. The minus sign in the latter two terms has non-trivial consequences for quantization.

Strictly speaking, the $A = 3, 4$ spinors are Dirac spinors only algebraically. Dynamically they are not Dirac spinors but, rather, what we call anti-Dirac spinors. The distinction between Dirac and anti-Dirac spinors is necessary, because action for free anti-Dirac spinors is the negative of the action for free Dirac spinor action and, hence, under the canonical quantization the assignment of creation and annihilation operators for $A = 3, 4$ spinors in (2.29) has to be reversed as compared to the standard Dirac spinor quantization assignment. This is the only way one can ensure non-negativity of contribution of $A = 3, 4$ spinors to the quantum Hamiltonian of the system described by (2.29) [28, 30, 33].

This concludes our reinterpretation of the standard gauge theory in terms of differential forms and bi-spinors. In the following section we will describe various forms of supersymmetric transformations that mix bosonic and fermionic difforms.

## 3. $U(1)$ Scalar Supersymmetry

We will now describe a realization of supersymmetry in the space that is a direct sum of spaces of commuting and anti-commuting difforms. Because the transformation parameters are Lorentz scalars, we will call it scalar supersymmetry (s-supersymmetry).

As we will see in this section, both in the simplest linear free field realization of s-supersymmetry as well as in the non-linear realization with interacting fields s-supersymmetry is broken by interactions. However, in our opinion this is actually an advantage, since for phenomenologically interesting asymptotically free theories s-supersymmetry is restored in the ultraviolet limit. Thus even s-supersymmetry of free field action contains two important features built-in from the beginning: It is explicitly broken at low energies, as is required by experimental evidence, and at the same time it is restored in the high energy limit, so that its benefits can apply for cancellation of the divergencies of the theory. S-supersymmetry for non-Abelian groups in examples of $SU(N) \times U(1)$ and $SU(N_3) \times SU(N_2) \times U(1)$ will be considered in the next section.

In order to make supersymmetry possible we have to make the same the number of formal commuting and anticommuting degrees of freedom. As for the standard supersymmetry some of the s-supersymmetric degrees of freedom will turn out to be non-dynamical. To match the bosonic and fermionic degrees of freedom we now first complexify difform $A_1$ in the Lagrangian in (2.21) to make it take values in the complexified form of the Lie algebra of $G$ and then promote it from a complex commuting 1-form $A_1$ to an arbitrary inhomogeneous commuting complex difform $A$, with the exception of complexified $A_1$ entering the definition of the gauged exterior derivative $d_{A_1}$. We no longer require that $G$ is semisimple. In fact we will see below that s-supersymmetry requires that $G$ contains at least one $U(1)$ factor. For the moment we will ignore the fact that gauge fields in the SM are real, i.e., take values in the real form of the Lie algebra of $G$, and will return to real gauge fields later in this section.

After complexification and augmentation with additional bosonic degrees of freedom we obtain the $U(1)$ Lagrangian that is the sum of the gauge invariant part $\mathcal{L}_{int}$, the gauge fixing part $\mathcal{L}_{gf}$, and the Faddeev–Popov ghost part $\mathcal{L}_{FP}$, which in the $U(1)$ case is trivial

$$\mathcal{L} = \mathcal{L}_{int} + \mathcal{L}_{gf} + \mathcal{L}_{FP},$$

$$\mathcal{L}_{int} = tr(\alpha\, d_{A_1} A, d_{A_1} A) + tr(\alpha\Phi, (d_{A_1} - \delta_{A_1})\Phi),$$

$$\mathcal{L}_{gf} = -\frac{1}{\xi_p} tr(\alpha\, \delta A_p, \delta A_p),$$

$$\mathcal{L}_{FP} = tr(\alpha\, \bar{c}, \delta d_{A_1} c),$$

(3.1)

where $A = \sum_{p=0}^{4} A_p$, $\Phi = \sum_{p=0}^{4} \Phi_p$ are complex valued. Note that we increased the number of Lagrange multipliers to a set of four $\xi_p$, $p = 1, \ldots, 4$, but omitted $\xi_0$, because for any 0-form $A_0$, $\delta A_0 = 0$. Thus, there are no gauge-fixing constraints on scalar field $A_0$. We have not increased the number of fermionic degrees of freedom. Thus the SM, when reformulated in terms of bi-spinors and s-supersymmetry, would not contain fermionic superpartners of any observed boson. As we will see below the additional supersymmetric partners of all of the observed fermions in such a reformulation can be packed into a single massless Higgs-like scalar field. Thus at the EW scale s-supersymmetry does not require any particle content additional to that of the SM, which, if s-supersymmetry is realized in Nature, would provide a natural explanation why one sees no evidence of the standard supersymmetry at LHC.

The equations of motion (2.23) are now augmented with equations for extra bosonic fields $A_{p \neq 1}$ in $A = \sum A_p$. We obtain

$$\delta_{A_1} d_{A_1} A + \frac{1}{\xi_1} d\delta A = J, \qquad (d_{A_1} - \delta_{A_1}) \Phi = 0,$$

$$(\alpha \delta A, \delta A) = 0, \qquad \delta d_{A_1} c = 0, \qquad \delta_{A_1} d\bar{c} = 0, \qquad (3.2)$$

$$(d_{A_1})^2 A_1 = 0, \qquad (\delta_{A_1})^2 A_1 = 0, \qquad \delta_{A_1} J = 0,$$

where the current diform $J = J_1$ is defined in (2.23). Since we preserved the special role of $A_1$, the gauge transformations reflect this. We become

$$\Phi \to \Omega_0 \Phi,$$
$$A_1 \to \Omega_0 A_1 \Omega_0^{-1} + \frac{i}{g} \Omega_0 (d - \delta) \Omega_0^{-1},$$
$$A_{p \neq 1} \to \Omega_0 A_{p \neq 1}.$$

The BRST transformations in (2.21) are also augmented with the gauge transformations for $A_{p \neq 1}$. We obtain the extended BRST in the form

$$\delta_\varepsilon \Phi = ig\,\varepsilon\,c\,\Phi, \qquad \delta_\varepsilon A_1 = \varepsilon d_{A_1} c, \qquad \delta_\varepsilon A_{p \neq 1} = ig\,\varepsilon\,c A_{p \neq 1},$$

(3.3)

$$\delta_\varepsilon c = i g\,\varepsilon c^2, \qquad \delta_\varepsilon \bar{c} = \frac{1}{\xi}\varepsilon\,\delta A_1.$$

Although $A_{p \neq 1}$ appears as extension of the gauge field part of the Lagrangian (3.1), $A_{p \neq 1}$ and $A_1$ play different roles in BRST transformations and in the Lagrangian. The reason is the special role 1-form $A_1$ plays in the definition of the covariant exterior derivative $d_{A_1} = d - ig A_1 \wedge$. The fact that $A_{p \neq 1}$ and $A_1$ are physically different can be also seen from analysis of [45, 46] of the physical content of theories based on $A_{p \neq 1}$ forms. We repeat the argument of [46] in a condensed form. First, since on $M_4$ 5-forms do not exist, the term with $A_4$ in $\mathcal{L}_{int}$ in (3.1) is absent. Therefore, from (3.2) $A_4$ is constant. Further, $A_3$ is also non-dynamical, while $A_2$ as a dynamical field can be reduced to $A_0$. $A_0$ itself is a complex scalar field, transforming in the fundamental representation, as we indicated in (3.3). Thus the total additional content of bosonic degree of freedom in Lagrangian (3.1) consists of two Higgs-like scalar fields. The fields $A_{p \neq 1}$ in (3.1) do not couple to fermions. We can force such coupling if instead of $(d_{A_1} - \delta_{A_1})$ in the fermionic Lagrangian we put in $(d_A - \delta_A)$, thus defining the covariant exterior derivative by $d_A = d - i g A \wedge$ instead of $d_{A_1} = d - i g A_1 \wedge$. This will induce a coupling between $A_{p \neq 1}$ and $\Phi$ that is similar to coupling of Higgs to fermions. We will not discuss this possibility here, since it would not affect our results on s-supersymmetry.

Just like BRST transformation (2.11) and bi-spinor BRST in (2.21) the extended bi-spinor BRST (3.3) is nilpotent, that is when applied to any field we obtain $\delta_\varepsilon^2 = 0$. For the additional bosonic fields this follows from $\varepsilon^2 = 0$. This property is essential for proving the unitarity of s-supersymmetric S-matrix in arbitrary $\xi$-gauge as the standard BRST for standard gauge theories. It should be as useful for proving the renormalizability of s-supersymmetric bi-spinor gauge theories as well. Although formal proof of renormalizability of s-supersymmetric bi-spinor gauge theories is outside the scope of this publication, the fact that coupling constant in bi-spinor gauge theory is dimensionless and the (extended) bi-spinor BRST is nilpotent provides strong indication of renormalizability of both bi-spinor gauge theories and s-supersymmetric bi-spinor gauge theories.

It should be mentioned as well that bi-spinor gauge theories and their s-supersymmetric extensions should possess the same as or better then anomaly cancellation properties. When four generations are present then bi-spinor gauge theory is anomaly free for any gauge group, because of pair-wise cancellation of anomalies among different generations. If bi-spinor gauge theory contains less the four generations then at least for the SM gauge group it is anomaly free, because anomaly cancellation in the SM takes place within each separate generation. Because bi-spinor SM differs from the SM only in the sign of some of the coupling constants, within generation cancellation of anomalies in bi-spinor SM follows from that of SM.

To proceed further we extract the quadratic part of the ghost-free part of Lagrangian (3.1) and obtain[5]

$$\mathcal{L}_0 = tr\left(\alpha(d-\delta)A,(d-\delta)A\right) - \lambda_p tr\left(\alpha\delta A_p, \delta A_p\right) + tr\left(\alpha\Phi,(d-\delta)\Phi\right), \quad (3.4)$$

where $\lambda_p = (1/\xi_p - 1/2)$, $p = 1,\ldots,4$. To obtain (3.4) we used $d^2 = \delta^2 = 0$ and, for convenience, combined $d$ and $\delta$ in the first term.

We can now describe the simplest realization of s-supersymmetry with Abelian gauge group $G = U(1)$. Non-Abelian groups of the SM will be considered in the following sections. In $\xi_p = 2$ gauge we obtain that free field $U(1)$ action is

$$S_0 = \int d^4x\, tr\left(\alpha(d-\delta)A,(d-\delta)A\right) + \int d^4x\, tr\left(\alpha\Phi,(d-\delta)\Phi\right). \quad (3.5)$$

In (3.5) $A$ is an arbitrary complex commuting diform, while $\Phi$ is an arbitrary complex anticommuting diform. We will call the pair $(A,\Phi)$ a complex multiplet. Since action (3.5) is real, the bosonic part of (3.5) splits in two non-interacting parts quadratic parts and thus, in effect, contains two unrelated sets of bosonic fields $A_R = \operatorname{Re} A$ and $A_I = \operatorname{Im} A$.

Our first main result is that action (3.5) is invariant with regard to transformations applied to $A$, $A^+$, $\Phi$, $\Phi^+$ considered as independent variables

$$\delta_\varepsilon A = \varepsilon\Phi, \quad \delta_\varepsilon A^+ = 0, \quad \delta_\varepsilon \Phi = 0, \quad \delta_\varepsilon \Phi^+ = -\varepsilon(d-\delta)A^+, \quad (3.6)$$

---

[5] In the mathematical literature operator $(d - \delta)$ is called the signature operator [57, 58].

$$\delta_{\varepsilon^*} A = 0, \quad \delta_{\varepsilon^*} A^+ = \varepsilon^* \Phi^+, \quad \delta_{\varepsilon^*} \Phi = -\varepsilon^* (d-\delta) A, \quad \delta_{\varepsilon^*} \Phi^+ = 0 \quad (3.7)$$

where $\varepsilon$, $\varepsilon^*$ are two complex-valued anticommuting transformation parameters, which are a Lorentz scalars. Parameters $\varepsilon$, $\varepsilon^*$ may be considered as independent, since action is invariant separately under (3.6) and (3.7). To derive invariance of (3.5) we used that under the conditions of Stokes' theorem, $\int_{M_4} df = \int_{\partial M_4} f$, operator $(d-\delta)$ is self-adjoint with respect to scalar product (A.14) and that $\Phi$ and $\varepsilon$ anticommute. Obviously, transformations (3.6-7) are a form of supersymmetry, for they mix the commuting bosonic and the anticommuting fermionic degrees of freedom, while leaving action (3.5) invariant. Also obvious is that (3.6-7) violate gauge symmetry, since transformation involve non gauge-covariant operator $(d-\delta)$.

Representing $A$, $\Phi$ in (3.6-7) in the $Z$-basis and using (A.18) and spinbein decomposition (A.30-32) of $\Psi(A), \Psi(\Phi)$ in terms of (anti-)Dirac spinors $\sigma^A(A)$, $\psi^A(\Phi)$, respectively, we obtain (3.6-7) in two different forms, where the non-zero variations are

$$\delta_\varepsilon \Psi(A) = \varepsilon \Psi(\Phi), \qquad \delta_{\varepsilon^*} \Psi(\Phi) = -\varepsilon^* (i\gamma^\mu \partial_\mu) \Psi(A),$$
$$\delta_\varepsilon \sigma^A(A) = \varepsilon \psi^A(\Phi), \qquad \delta_{\varepsilon^*} \psi^A(\Phi) = -\varepsilon^* (i\gamma^\mu \partial_\mu) \sigma^A(A). \quad (3.8)$$

Notice that the last two transformations in (3.8) do not mix generation index $A$. Considering that multiplication with non-degenerate matrices $\sigma^A{}_\alpha(A)$, $\psi^A{}_\alpha(\Phi)$ preserves matrix rank, we obtain that the constraints $\det \Psi^{ab}(A)=0$, $\det \Psi^{ab}(\Phi)=0$, where determinant is taken over Lorentz matrix indices, are consistent with scalar supersymmetry and thus (3.8) are also supersymmetry transformations for bi-spinors $\Psi(A), \Psi(\Phi)$ containing less then four generations of Dirac-(anti)-Dirac spinors $\psi^A(\Phi)$ in their spinbein decompositions, provided that we use a degenerate spinbein with appropriate rank less than four, for example, spinbein (A.33) of rank three. Thus, if necessary, the fourth generation $\psi^4(\Phi)$ can be made to completely decouple from the other three using a degenerate spinbein and can be effectively put to zero.

From (3.6-7) we obtain the commutators of two s-supersymmetry transformations

$$[\delta_\varepsilon, \delta_{\varepsilon^*}] A = \varepsilon^* \varepsilon (d-\delta) A, \qquad [\delta_\varepsilon, \delta_{\varepsilon^*}] \Phi = \varepsilon^* \varepsilon (d-\delta) \Phi,$$
$$[\delta_\varepsilon, \delta_{\varepsilon^*}] A^+ = \varepsilon \varepsilon^* (d-\delta) A^+, \qquad [\delta_\varepsilon, \delta_{\varepsilon^*}] \Phi^+ = \varepsilon \varepsilon^* (d-\delta) \Phi^+. \quad (3.9)$$

Defining two charges $Q_\varepsilon$, $Q_{\varepsilon^*}$ by $\delta_\varepsilon = \varepsilon Q_\varepsilon$, $\delta_{\varepsilon^*} = \varepsilon^* Q_{\varepsilon^*}$ we obtain the anticommutator algebra of the corresponding s-supersymmetry charges acting in the space of ($A$, $\Phi$, $A^+, \Phi^+$)

$$\{Q_\varepsilon, Q_\varepsilon\}=0, \qquad \{Q_{\varepsilon^*}, Q_{\varepsilon^*}\}=0, \qquad \{Q_\varepsilon, Q_{\varepsilon^*}\} = (d-\delta). \quad (3.10)$$

In the Z-basis the two charges take a simple form. First we note that in (3.9) the global s-supersymmetry is actually also a local s-supersymmetry. That is if in (3.9) we assume that $\varepsilon = \varepsilon(x)$ then action (3.5) is still in invariant with respect to (3.9). This is the result of the existence of conserved currents and associated charges defined by

$$J_\varepsilon^\mu = tr\left[\overline{\overline{\Psi}}(A)\tilde{\partial}\gamma^\mu \Psi(\Phi)\right], \qquad \partial_\mu J_\varepsilon^\mu = 0,$$

$$Q_\varepsilon = \int d^3x\, tr\left[\overline{\overline{\Psi}}(A)\tilde{\partial}\gamma^0 \Psi(\Phi)\right].$$

(3.11)

$$J_{\varepsilon^*}^\mu = tr\left[\overline{\overline{\Psi}}(\Phi)\gamma^\mu \partial \Psi(A)\right], \qquad \partial_\mu J_{\varepsilon^*}^\mu = 0,$$

$$Q_{\varepsilon^*} = \int d^3x\, tr\left[\overline{\overline{\Psi}}(\Phi)\gamma^0 \partial \Psi(A)\right],$$

(3.12)

where we defined currents through action variation as

$$\delta S = \delta S(\varepsilon = const) + \int d^4x \left(\partial_\mu \varepsilon\, J_\varepsilon^\mu + J_{\varepsilon^*}^\mu \partial_\mu \varepsilon^*\right).$$

Conservation of the currents follows from Noether's theorem and can be verified using equations of motion derived from (3.5) and converted to the Z-basis. Expressions (3.11-12) for the currents and the charges give an example of how much simpler expressions for dynamical quantities could look when both bosonic and the fermionic difforms are expressed in the Z-basis. The analogous expressions for the currents and charges expressed in the c-basis are rather unwieldy.

Expressions (3.9-10) should be compared with the commutator of two transformations of the standard $N = 1$ supersymmetry with no central charges and the anticommutator of corresponding charges on $M_4$. This is given by

$$[\delta, \overline{\delta}] = 2\overline{\theta}\gamma^\mu P_\mu \theta, \qquad \{Q, \overline{Q}\} = 2\gamma^\mu P_\mu,$$

(3.13)

where $P_\mu = i\partial_\mu$ is the translation operator and $\theta, \overline{\theta}$, are two infinitesimal Grassmann parameters transforming as Dirac spinors. We observe that the standard and s-supersymmetry are related via the transformation that maps difform $(d - \delta)\Phi$ into the set of its coefficients $i\partial \Psi(\Phi)$ in the Z-basis. Therefore, on $M_4$ we can consider s-supersymmetry as a coordinate-free version of the standard $N = 1$ supersymmetry.

It is important to stress that relations (3.9-10) are valid on any smooth manifold, while relations of type (3.12) can only be defined on manifolds that admit spin structure. Therefore, in general, one cannot speak of equivalence of scalar and standard supersymmetry. Instead it is more accurate to say that s-supersymmetry is a notion inequivalent to that of the standard supersymmetry and coincides with it only for special cases. We will see other examples of possible inequivalence below and postpone more detailed analysis of the correspondence till another occasion. However, we should note that supersymmetric algebra similar to algebra (3.10) may appear as subalgebra of the algebra of twisted extended Euclidean lattice supersymmetry acting on lattice bi-spinors, which in the lattice literature are usually referred to as Dirac-Kähler spinors. It appears as scalar term of representation of the extended supersymmetry algebra in the Z-basis [16].

In this Letter we consider only massless fields. We will discuss briefly how to introduce a mass parameter for complex multiplet $(A, \Phi)$. The simplest but not the only way is to substitute in s-supersymmetric transformation (3.6-7) and action (3.5) massless operator $(d - \delta)$ with $(d - \delta - m)$, where $m$ is a mass parameter. It is easy to see that such substitution) leaves modified action (3.5) unchanged. Instead of (3.10), after substitution we obtain

$$\{Q_\varepsilon, Q_\varepsilon\} = 0, \qquad \{Q_{\varepsilon^*}, Q_{\varepsilon^*}\} = 0, \qquad \{Q_\varepsilon, Q_{\varepsilon^*}\} = (d - \delta - m). \quad (3.14)$$

In the Z-basis this becomes

$$\{Q, Q\} = 0, \qquad \{\overline{Q}, \overline{Q}\} = 0, \qquad \{Q, \overline{Q}\} = 2\gamma^\mu P_\mu - 2m. \quad (3.15)$$

Therefore, introduction of mass parameter by substitution of $(d - \delta)$ with $(d - \delta - m)$ results in the central charge $-2m$ in the s-supersymmetry algebra. Such substitution always leads to degenerate mass spectrum of the (anti-)Dirac spinors. It turns out that analysis of the most general explicit mass terms via operators of dimension three in bi-spinor gauge theory is somewhat non-trivial and, therefore, is outside the scope of this Letter. We refer the reader to [32, 33] for details.

One interesting fact about the standard Euclidean lattice supersymmetry is that hypercubic lattice that is typically used for realizations of the standard supersymmetry allows for natural appearance of extended supersymmetry, which is needed to construct twisted realizations of the standard supersymmetry, the generators of which can be enumerated by the big diagonals of the basic lattice hypercube. Whether one can introduce extended s-supersymmetry on a smooth manifold is an open question, the answer to which relies on the full analysis of all possible s-supersymmetric algebras. Also the exact connection between s-supersymmetry and supersymmetry of twisted Dirac-Kähler fields on the lattice remains to be explored.

We should also comment briefly on the long-standing apparent puzzle about representation of fermions by a collection of anticommuting antisymmetric tensors. From our point of view, the puzzle is resolved through the spinbein decomposition (A.32) of bi-spinors. In a general spinbein gauge a bi-spinor is completely analogous to a collection of anti-symmetric tensors. Both are coefficients of expansion of a difform in a particular basis. However, for physical spinbein gauges, where spinbeins must be constant [28, 30], so that they can transfer all their dynamical degrees of freedom to algebraic Dirac spinors, one can consider space-time as if it carries spin. Just like constant energy, on $M_4$ this spin is not detectable, because of constancy of spinbein. Thus, from the point of view of physical (anti-)Dirac spinors contained in bi-spinors, the vacuum state in bi-spinor gauge theories may be considered as a spin state. Combined with spin of the (anti-)Dirac spinors the vacuum state turns spin of otherwise spin one-half states into spin of integer spin states represented by anti-symmetric tensors.

Since a particular $\xi$-gauge with $\xi_p = 2$ is required for realization of s-supersymmetry, we have to ask ourselves whether such symmetry is physically acceptable. The answer is yes, it is physically acceptable. First, recall from the example of chiral symmetry broken by mass terms that symmetries of only a part of the Lagrangian can have important physical consequences. Consider also the quantum Dirac spinor gauge theory described by quantum gauge-fixed Lagrangian (2.3). There, gauge symmetry is also symmetry of only a part of the Lagrangian. Similarly, in the quantum gauge-fixed Lagrangian in (2.21) of the bi-spinor gauge theory only a part of the Lagrangian is gauge invariant. Of course, only gauge independent consequences of the presence of symmetry broken or not can be physical. This does not prevent, however, the appearance at some intermediate stages of gauge dependent expressions.

It is well-known that in different gauges different desirable features of a gauge theory become more pronounced. We can certainly consider unitarity of $S$-matrix as manifestation of symmetry, a consequence of requirement of conservation of probability. However, this symmetry is transparent only in the unitary gauge. The same applies to covariance of a gauge theory. Only in $\xi$-gauges the covariance is manifest but not in the unitary gauge. In the same vein we can consider $\xi = 2$ gauge of bi-spinor gauge theory as the gauge where s-supersymmetry is manifest or alternatively we can consider it as symmetry of a part of the Lagrangian by analogy with broken chiral symmetry. In either case its effects can be judged only on gauge invariant and measurable quantities. In gauges other then $\xi = 2$ it is not manifest, but what matters of course is whether $S$-matrix amplitudes in some way reflect the presence of partial s-supersymmetry. We will consider this question in more detail elsewhere.

Another interesting aspect of s-supersymmetry is that for asymptotically free theories, the theories that are actually of physical interest, approximate s-supersymmetry becomes exact in the ultraviolet limit. Thus asymptotically free bi-spinor gauge theory possesses the two necessary ingredients for solution of hierarchy problem. At low energy s-supersymmetry is broken, while at high energy it is restored, so that one should expect all the benefits of cancellation of divergencies that come from the standard supersymmetry, where one has to break supersymmetry softly, a non-trivial task in itself. We note recent work in [47], where the authors explore a scenario were broken standard supersymmetry is restored in the ultraviolet limit by quantum effects.

The requirement that bosonic difform $A$ in (3.1) is complex seems to be physically unacceptable[6], since at least for compact gauge groups it may violate unitarity of the representation of the symmetry. Hence, we consider the simplest realization of s-supersymmetry only as an illustrative example. We will now describe a realization of s-supersymmetry with physical gauge fields. To start with, note that, as can be seen from the definition of the chiral star operator in (A.12), unlike in the Euclidean space-time, in the Minkowski space-time there are no real chiral bi-spinors. Therefore, to provide physically acceptable realizations of s-supersymmetry we have to restrict ourselves to real-valued bosonic difforms but we cannot use real-valued fermionic difforms. This means that to match the degrees of freedom we need to reduce their number for fermions by half, while keeping them complex-valued.

The simplest way to do this is to use chiral fermionic difforms we defined in the Appendix with the use of chiral star operator (A.12)[7]. In addition, for chosen chirality – we will concentrate on left chiral difforms only - we have to make use of two operators: the left conversion operator $L_A$, which maps the space of real difforms into

---

[6] Complex connection difforms and coupling constants do occur in theories of gravity.
[7] Another way to halve fermionic degrees of freedom would be to use Majorana difforms.

the space of left chiral complex difforms, and the left conversion operator $L_\Phi$, which acts in the opposite direction. The most obvious left conversion operators are parameterized by a real parameter $\mu$, $\mu \neq 0$, $\mu \neq \infty$, and are given by

$$L_A : A \to \Phi_L, \qquad L_A = \sqrt{2} P_L \left(1 + i\mu^{-1}(d-\delta)\right) P_+, \qquad (3.16)$$

$$L_\Phi : \Phi_L \to A, \qquad L_\Phi = \sqrt{2} P_+ \left(1 - i\mu^{-1}(d-\delta)\right) P_L, \qquad (3.17)$$

where the left chiral difforms $\Phi_L$ are defined by (A.28-29) and

$$P_+ A \equiv A_R = (1/2)(A + A^*) = (1/2)(1 + \overline{C})A, \qquad \overline{C} A \equiv A^*, \qquad (3.18)$$

$$P_- A \equiv A_I = (1/2i)(A - A^*) = (1/2i)(1 - \overline{C})A \qquad (3.19)$$

are projectors on the real and imaginary parts of a complex-valued difform $A$. For difforms with values in Lie algebras complex conjugation is replaced with the hermitean conjugation. Note that operator $i(d-\delta)$ in (3.16-17) is needed to make (3.16-17) nontrivial. Its application converts real difforms into complex difforms and left chiral difforms into right chiral difforms. Parameter $\mu$ of dimension of mass is needed to compensate for dimension of $(d-\delta)$. The right conversion operators $R_A$, $R_\Phi$ are obtained from (3.13-14) by $P_L \to P_R$. It follows from (3.16-19) that

$$L_A L_\Phi : \Phi_L \to \Phi_L, \qquad L_A L_\Phi = \left(1 + \mu^{-2}(d-\delta)^2\right) P_L, \qquad (3.20)$$

$$L_\Phi L_A : A \to A, \qquad L_\Phi L_A = \left(1 + \mu^{-2}(d-\delta)^2\right) P_+, \qquad (3.21)$$

$$R_\Phi^+ = L_A, \qquad L_\Phi^+ = R_A, \qquad (3.22)$$

where Hermitean conjugation in (3.22) is with respect to scalar product (A.14).

We can now describe s-supersymmetry realization for $U(1)$ $\xi = 2$ action for free real Abelian massless bosonic fields $A_R$ and free massless chiral bi-spinor fermionic fields $\Phi_L$. We will call the pair $(A_R, \Phi_L)$ a chiral multiplet. Action (3.5) now is

$$S = \int d^4 x\, tr\left(\alpha (d-\delta) A_R, (d-\delta) A_R\right) + \int d^4 x\, tr\left(\alpha \Phi_L, (d-\delta) \Phi_L\right), \qquad (3.23)$$

which, using projection operators (3.18-19, A.28), we can rewrite (3.23) in an equivalent form

$$S = \int d^4 x\, tr\left(\alpha (d-\delta) P_+ A, (d-\delta) P_+ A\right) + \int d^4 x\, tr\left(\alpha P_L \Phi, (d-\delta) P_L \Phi\right). \qquad (3.24)$$

In (3.24) difform $A$ is an arbitrary complex commuting difform, while $\Phi$ is an arbitrary complex anticommuting difform. It is easy to see with the use of (3.22) that action (3.23) is invariant with regard to the *c*- or *Z*-basis infinitesimal transformations

$$\delta_\varepsilon A_R = i\varepsilon L_\Phi \Phi_L, \qquad\qquad \delta_\varepsilon \Phi_L = i\varepsilon L_A (d-\delta) A_R,$$

$$\delta_\varepsilon \Psi(A_R) = i\varepsilon \hat{L}_\Phi \Psi(\Phi_L), \qquad \delta_\varepsilon \overline{\Psi}(A_R) = i\varepsilon \overline{\Psi}(\Phi_L)\overline{\hat{\tilde{L}}}_\Phi, \qquad (3.25)$$

$$\delta_\varepsilon \Psi(\Phi_L) = i\varepsilon \hat{L}_A (i\partial)\Psi(A_R), \qquad \delta_\varepsilon \overline{\Psi}(\Phi_L) = i\varepsilon \overline{\Psi}(A_R)(i\overleftarrow{\partial})\overline{\hat{\tilde{L}}}_A,$$

where $\overline{\hat{\tilde{L}}}_A = \gamma^0 \hat{\tilde{L}}_A^+ \gamma^0$, $\overline{\hat{\tilde{L}}}_\Phi = \gamma^0 \hat{\tilde{L}}_\Phi^+ \gamma^0$ are the Z-basis analogs of $L_A^+, L_\Phi^+$ adjoint to $L_A, L_\Phi$ with respect to (A.14) except that all derivatives act from the right and where arrow over $\overline{\hat{\tilde{L}}}$ indicates direction of action of derivatives. $\varepsilon$ is a single real anticommuting transformation parameter: $\varepsilon^* = \varepsilon$. It must be real so that $A$ can be real, as can be seen from

$$(\delta_\varepsilon A_R)^+ = (i\varepsilon L_\Phi \Phi_L)^+ = -i(L_\Phi \Phi_L)^+ \varepsilon^* = i\varepsilon L_\Phi \Phi_L = \delta_\varepsilon A_R, \qquad (3.26)$$

since by construction $(L_\Phi \Phi_L)^+ = L_\Phi \Phi_L$.

Invariance of (3.23) under transformation (3.25) is our second main result. Note that s-supersymmetric transformation of chiral supermultiplet involves a single parameter and, hence, a single charge $Q$, with trivial algebra $\{Q,Q\} = 2Q^2 = 0$. The conserved current (the transformation parameter is taken as $i\varepsilon$) and the charge are obtained from Noether's theorem

$$J_\varepsilon^\mu = tr\left[ \overline{\Psi}(\Phi_L)\overline{\hat{\tilde{L}}}_\Phi \gamma^\mu (i\overleftarrow{\partial})\Psi(A_R) - \overline{\Psi}(A_R)(-i\overleftarrow{\partial})\gamma^\mu \hat{L}_\Phi \Psi(\Phi_L) \right], \qquad \partial_\mu J_\varepsilon^\mu = 0, \quad (3.27)$$

$$Q_\varepsilon = \int d^3x J_\varepsilon^0. \qquad (3.28)$$

As expected, if we set $\hat{L}_\Phi = 1$ then $J_\varepsilon^\mu$ reduces to the sum of two currents in (3.11-12). Conservation of current $J_\varepsilon^\mu$ follows from equations of motion and from $\partial \hat{L}_\Phi = \hat{R}_\Phi \partial$. We conclude, that because of conservation of current $J_\varepsilon^\mu$ globally defined s-supersymmetry (3.25) with chiral multiplet $(A_R, \Phi_L)$ is actually also a local s-supersymmetry. Unlike the standard supersymmetry no additional fields were required to make the transition from the global to the local case.

We will now consider s-supersymmetry for interacting Lagrangian for chiral multiplet. At the present we are able to give examples of non-linear on-shell realizations of s-supersymmetry. Non-linearity of interacting s-supersymmetry makes it less interesting as symmetry of quantum bi-spinor gauge theory. From the point of view of phenomenological applications, the most interesting is the linear partial s-supersymmetry: it has all the desired features of supersymmetry, that is, it is broken at low energies and is exact in the ultraviolet limit for asymptotically free theories.

Whether non-linear s-supersymmetry of interacting action can be extended to a linear realization is an open question.

Consider the gauged version of action (3.23) with chiral multiplet $(A_R, \Phi_L)$. In its symmetric form for fermionic action it is

$$S = \int d^4 x \, tr(\alpha \nabla_{A_1} A_R, \nabla_{A_1} A_R) + \frac{1}{2}\int d^4 x (\alpha \Phi_L, \nabla_{A_1} \Phi_L) + \frac{1}{2}\int d^4 x (\alpha \nabla_{A_1} \Phi_L, \Phi_L), \quad (3.29)$$

where we denoted $\nabla_{A_1} = d_{A_1} - \delta_{A_1}$. We now modify the conversion operators (3.16-17) to include gauged versions of the exterior derivative and covariant divergence. We define their actions both in the c- and the Z-bases, with the latter distinguished from the former by a hat over the operator

$$L_A : A_R \to \Phi_L, \qquad L_A = \sqrt{2} P_L (1 + i\mu^{-1}\nabla_{A_1}) P_+, \quad (3.30)$$

$$\hat{L}_A : \Psi(A_R) \to \Psi(\Phi_L), \qquad \hat{L}_A = \sqrt{2} \hat{P}_L (1 - \mu^{-1}\partial_{A_1}) \hat{P}_+,$$

$$\hat{L}_A \Psi(A_R) \equiv \Psi(L_A A_R), \quad (3.31)$$

$$L_\Phi : \Phi_L \to A_R, \qquad L_\Phi = \sqrt{2} P_+ (1 - i\mu^{-1}\nabla_{A_1}) P_L, \quad (3.32)$$

$$\hat{L}_\Phi : \Psi(\Phi_L) \to \Psi(A), \qquad \hat{L}_\Phi = \sqrt{2}\hat{P}_+ (1 + \mu^{-1}\partial_{A_1}) \hat{P}_L,$$

$$\hat{L}_\Phi \Psi(\Phi_L) \equiv \Psi(L_\Phi \Phi_L), \quad (3.33)$$

where in the Z-basis

$$\hat{P}_{L,R} = (1/2)(1 \mp \gamma^5), \quad (3.34)$$

$$i\partial_{A_1} = i\partial + g\, \Psi(A) = i\partial + \frac{g}{4} A_1, \quad (3.35)$$

$$\hat{P}_\pm \Psi = \frac{1}{2}(\Psi \pm \beta\gamma^0 \overline{C}\Psi\gamma^0) \equiv \frac{1}{2}(\Psi \pm \beta \overline{\overline{\Psi}}). \quad (3.36)$$

The two gauged conversion operators satisfy relations analogous to (3.20-22)

$$\hat{L}_A \hat{L}_\Phi : \Psi(\Phi_L) \to \Psi(\Phi_L), \qquad \hat{L}_A \hat{L}_\Phi = (1 + \mu^{-2}\partial_{A_1}^2)\hat{P}_L, \quad (3.37)$$

$$\hat{L}_\Phi \hat{L}_A : \Psi(A_R) \to \Psi(A_R), \qquad \hat{L}_\Phi \hat{L}_A = (1 + \mu^{-2}\partial_{A_1}^2)\hat{P}_+, \quad (3.38)$$

$$\overline{\overline{\hat{R}}}_\Phi = \hat{L}_A, \qquad \overline{\overline{\hat{R}}}_A = \hat{L}_\Phi, \quad (3.39)$$

where in (3.39) hermitean conjugation is with respect (A.14).

We will now consider non-linear transformations of chiral multiplet $(A_R, \Phi_L)$ and their effect on action (3.29). For convenience we will work in the Z-basis. The variations in c- and Z-bases are given by

$$\delta_\varepsilon A_R = i\varepsilon L_\Phi \Phi_L, \qquad \delta_\varepsilon \Phi_L = i\varepsilon L_A \nabla_{A_1} A_R,$$

$$\delta_\varepsilon \Psi_A = i\varepsilon \hat{L}_\Phi \Psi_\Phi, \qquad \delta_\varepsilon \Psi_\Phi = i\varepsilon \hat{L}_A (i\vec{\partial}_{A_1}) \Psi_A,$$

$$\delta_\varepsilon \overline{\overline{\Psi}}_A = i\varepsilon \overline{\overline{\Psi}}_\Phi \overline{\overline{\hat{L}}}, \qquad \delta_\varepsilon \overline{\overline{\Psi}}_\Phi = -i\varepsilon \overline{\overline{\Psi}} (-i\overleftarrow{\partial}_{-A_1}) \overline{\overline{\hat{L}}}_A,$$

(3.40)

$$\delta_\varepsilon (i\vec{\partial}_{A_1}) = \delta_\varepsilon (-i\vec{\partial}_{-A_1}) = \frac{ig\varepsilon}{2\sqrt{2}} \gamma^\mu \tilde{\Phi}_\mu, \qquad \tilde{\Phi}_\mu \equiv \mathrm{Re}\, \Phi_\mu.$$

Variation of action (3.23) consists of two parts: the "free" part, which is identical with the variation of free action where $\nabla_{A_1}$ is substituted in place of $(d - \delta)$, and the residual part that involves only variation of $A_1$ in $\nabla_{A_1}$

$$\delta_\varepsilon S = \delta_\varepsilon^{free} S + \delta_\varepsilon^{res} S,$$

(3.41)

The first part, $\delta_\varepsilon^{free} S$, vanishes because of (3.39) and self-adjointness of $\nabla_{A_1}$. For the residual part we obtain

$$\delta_\varepsilon^{res} S = \delta_\varepsilon S_b + \delta_\varepsilon S_f,$$

(3.42)

where in the Z-basis $\delta_\varepsilon S_b$ and $\delta_\varepsilon S_f$ are given by

$$\delta_\varepsilon S_b = \frac{ig\varepsilon}{2\sqrt{2}} \int d^4x \left( tr\left[ \overline{\overline{\Psi}}_A \gamma^\mu \tilde{\Phi}_\mu \vec{\partial}_{A_1} \Psi_A - \overline{\overline{\Psi}}_A \overleftarrow{\partial}_{-A_1} \gamma^\mu \tilde{\Phi}_\mu \Psi_A \right] \right),$$

$$\delta_\varepsilon S_f = -\frac{ig\varepsilon}{2\sqrt{2}} \int d^4x \left( tr\left[ \overline{\overline{\Psi}}_\Phi \gamma^\mu \tilde{\Phi}_\mu \Psi_\Phi \right] \right),$$

(3.43)

where $\Psi_A \equiv \Psi(A_R)$, $\Psi_\Phi \equiv \Psi(\Phi_L)$ and we used

$$\left( i\overleftarrow{\overline{\partial}}_{A_1} \right) \overline{\overline{\Psi}}_A = \overline{\overline{\Psi}}_A \left( -i\vec{\partial}_{-A_1^+} \right),$$

$$(A_R)^+ = A_R,$$

$$(\hat{L}_\Phi \Psi_\Phi)_1 = \sqrt{2} \hat{P}_+ (\Psi_\Phi)_1 = \frac{1}{2\sqrt{2}} \gamma^\mu \tilde{\Phi}_\mu,$$

(3.44)

$$\overline{\overline{\Psi}}(A_R) = \beta \Psi(A_R),$$

$$\overline{\overline{\Psi}}((L_\Phi \Phi_L)_1) = \beta \Psi((L_\Phi \Phi_L)_1) = \Psi((L_\Phi \Phi_L)_1) = \frac{1}{2\sqrt{2}} \gamma^\mu \tilde{\Phi}_\mu.$$

Using (3.40, 3.44), $tr(ABC) = tr(CBA)$, $\hat{P}_L \Phi_L = \Phi_L = 1/2(1-\gamma^5)\Phi$, and anti-commutativity of $\Phi$, we obtain that (3.42-43) can be written as

$$\delta_\varepsilon^{res} S = \frac{ig\varepsilon}{2\sqrt{2}} \int d^4x\, K^\mu \tilde{\Phi}_\mu,$$

$$K^\mu = tr\left[\overline{\overline{\Psi}}_A \gamma^\mu (i\vec{\partial}_{A_1})\Psi_A + \overline{\overline{\Psi}}_A (-i\vec{\partial}_{-A_1})\gamma^\mu \Psi_A + \overline{\overline{\Psi}}_\Phi \gamma^\mu \Psi_\Phi \right].$$
(3.45)

We see that $\delta_\varepsilon^{res} S$ vanishes if a single constraint is satisfied

$$K^\mu \tilde{\Phi}_\mu = 0.$$
(3.46)

To satisfy (3.46), for example, we can assume that $\tilde{\Phi}_\mu$ is arbitrary and impose stronger sufficient conditions on bosonic and fermionic parts of (3.46)

$$tr\left[\left((i\vec{\partial}_{A_1})\Psi_A \overline{\overline{\Psi}}_A + \Psi_A \overline{\overline{\Psi}}_A (-i\vec{\partial}_{-A_1})\right)\gamma^\mu \right] = 0,$$

$$tr\left[\Psi_\Phi \overline{\overline{\Psi}}_\Phi \gamma^\mu \right] = 0, \quad \mu = 0,\cdots,3.$$
(3.47)

Rewriting the first condition in (3.47) as

$$tr\left[\left(\gamma^\lambda (i\partial_\lambda \Psi_A + g A_\lambda \Psi_A)\overline{\overline{\Psi}}_A + \Psi_A (-i\partial_\lambda \overline{\overline{\Psi}}_A + g \overline{\overline{\Psi}}_A A_\lambda)\gamma^\lambda \right)\gamma^\mu \right] = 0,$$

we can reduce the bosonic constraint to two independent conditions valid for any value of coupling constant $g$

$$tr\left[\Psi_A \overline{\overline{\Psi}}_A\right] = 0,$$

$$tr\left[\left(\partial_\lambda \Psi_A \overline{\overline{\Psi}}_A - \Psi_A \partial_\lambda \overline{\overline{\Psi}}_A \right)\gamma^{|\mu} \gamma^{\lambda|}\right] = 0,$$

$$tr\left[\Psi_\Phi \overline{\overline{\Psi}}_\Phi \gamma^\mu \right] = 0, \quad \mu = 0,\cdots,3,$$
(3.48)

where we have to keep in mind that, because of bosonic difform reality and fermionic difform chirality, we have

$$\overline{\overline{\Psi}}_A = \beta \Psi_A \;, \qquad \beta \Psi_A = \sum_p \gamma^{\langle \mu_p} \cdots \gamma^{\mu_1 \rangle} A_{|\mu_1 \ldots \mu_p|} \;,$$

$$\hat{P}_L \Psi_\Phi = \Psi_\Phi \;. \tag{3.49}$$

Let us consider the bosonic and fermionic conditions in more detail. Given an expansion of $\Psi_A$ in terms of 16 independent coefficients in orthogonal basis $\gamma^{|\mu_1|} \cdots \gamma^{|\mu_p|}$ of powers of $\gamma$-matrices, we can write out matrix $\Psi_A \overline{\overline{\Psi}}_A$ in the same basis as

$$\Psi_A \overline{\overline{\Psi}}_A = \sum_p \gamma^{|\mu_1|} \cdots \gamma^{|\mu_p|} F(A_R)_{|\mu_1 \ldots \mu_p|} \;,$$

$$tr\left[\Psi_A \overline{\overline{\Psi}}_A\right] = \sum_p (A_R)_{|\mu_1 \ldots \mu_p|} (A_R)^{|\mu_1 \ldots \mu_p|} \;. \tag{3.50}$$

Just like $\Psi_A$ matrix $\Psi_A \overline{\overline{\Psi}}_A$ also has 16 real independent components $F_{|\mu_1 \ldots \mu_p|}$. We now see from orthogonality of $\gamma^{|\mu_1|} \cdots \gamma^{|\mu_p|}$ that the meaning of (3.48) is that certain terms in the expansion (3.50) for $\Psi_A \overline{\overline{\Psi}}_A$, $\partial_\mu \Psi_A \overline{\overline{\Psi}}_A$ and analogous expansion for $\Psi_\Phi \overline{\overline{\Psi}}_\Phi$ are missing.

The first condition in (3.49) can then be considered as algebraic relation that sets $(A_R)_{0123}$ to a certain value up to a sign. Since when gauge is fixed $(A_R)_{0123}$ is not present in the bosonic gauge-fixed Lagrangian $tr(\alpha \nabla_{A_1} A_R, \nabla_{A_1} A_R)$, $(A_R)_{0123}$ does not enter equations of motion for gauge difform and, hence, can be freely set via (3.48, 3.50). What then remains of the bosonic part of (3.48) is

$$tr\left[\left(\partial_\lambda \Psi_A \overline{\overline{\Psi}}_A - \Psi_A \partial_\lambda \overline{\overline{\Psi}}_A \right) \gamma^{|\mu} \gamma^{\lambda|}\right] = 0 \;. \tag{3.51}$$

These are four real conditions on $A_R$ that can be considered as the conditions on the four independent components $\tilde{A}_\lambda$ of $(A_R)_{\mu\nu\sigma} = \varepsilon_{\mu\nu\sigma}{}^\lambda \tilde{A}_\lambda$. Since $\tilde{A}_\lambda$ are non-dynamical quantities [46], we can consider (3.51) as first order differential constraints that fix it. Thus, in the end we obtain that imposing on-shell s-supersymmetry on full interacting action (3.29) with a chiral multiplet and constraints (3.48) fixes the non-dynamical degrees of freedom of $A_R$ in terms of its dynamical degrees of freedom. The latter consist of gauge field $A = (A_R)_1$ and two scalar fields $H_0 = (A_R)_0$ and $H_1$ extracted from $(A_R)_2$, both of which transform in the fundamental representation and thus could be combined into a single complex scalar field. This is exactly the bosonic spin content of the SM.

Let us now turn to fermionic constraint. Using spinbein decomposition with spinbein $\eta^A$: $\Psi(A_R) = \psi^A \overline{\overline{\eta}}^A$, it can be written in four and two component spinor notation as

$$\overline{\overline{\psi}}_L^{\ A} \gamma^\mu \psi_L^{\ A} = 0, \qquad \psi_L^{\ A} = \begin{bmatrix} \chi^A \\ 0 \end{bmatrix}, \qquad \chi^A = \begin{bmatrix} \chi_1^A \\ \chi_2^A \end{bmatrix},$$

$$\Gamma^{AB} \chi^{A^+} \sigma^\mu \chi^B = 0, \qquad \gamma^\mu = \begin{pmatrix} 0 & \sigma^\mu \\ \overline{\sigma}_\mu & 0 \end{pmatrix}, \qquad \gamma^5 = \begin{pmatrix} -1 & 0 \\ 0 & 1 \end{pmatrix},$$

(3.52)

where $\Gamma^{AB} = diag(1, 1, -1 -1)$ and $\psi_L^{\ A}$, $\chi^A$ are four generations of left-handed Dirac-(anti)-Dirac spinors and two component Weil spinors, respectively. The four equations (3.52) allow us to express up to a common phase two complex parameters $\chi_k^A$, $k = 1,2$, of two component spinor $\chi^A$ of one generation, say with $A = 4$ in terms of the remaining three.

Since bi-spinor gauge theories are globally $U(2,2)$ invariant one can always rotate the four (anti)-Dirac spinors in the generation space so that (3.52) reduces to $\overline{\overline{\psi}}_L^{\ 4} \gamma^\mu \psi_L^{\ 4} = 0$. This constraint, in turn, would be identically satisfied if we choose a degenerate spinbein (A.33) that cuts the fourth generation from dynamics. We conclude, that requiring s-supersymmetry with chiral multiplet necessarily leads to three generations of (anti)-Dirac spinors. Thus, the consequence of s-supersymmetry on the particle spectrum of four-dimensional bi-spinor gauge theory is that it is identical to particle spectrum of massless SM.

We will now outline how to incorporate constraint (3.46) into the dynamics with the use of the Lagrange multiplier method. Details of the derivation together with hamiltonian analysis of the algebra of constraints (3.48) within the framework of Dirac constraint method will be given elsewhere. To use Lagrange multiplier method we modify action (3.29) by adding an additional term to obtain

$$S_\rho = \int d^4 x \, tr(\alpha \nabla_{A_1} A_R, \nabla_{A_1} A_R) + \int d^4 x (\alpha \Phi_L, \nabla_{A_1} \Phi_L) + \rho \int d^4 x (\alpha K, \tilde{\Phi}), \qquad (3.53)$$

where $\rho$ is the Lagrange multiplier and $K, \tilde{\Phi}$ are 1-forms $K = K_\mu dx^\mu$, $\tilde{\Phi} = \tilde{\Phi}_\mu dx^\mu$. We now have to verify that the last term in (3.53) is itself on-shell s-supersymmetric. That is that

$$\int d^4 x \, \delta_\varepsilon \, tr[K^\mu \tilde{\Phi}_\mu] = 0, \qquad (3.54)$$

where from (3.45)

$$K^\mu = tr\left[ \overline{\Psi}_A \gamma^\mu (i\vec{\partial}_{A_1}) \Psi_A + \overline{\Psi}_A (-i\vec{\partial}_{-A_1}) \gamma^\mu \Psi_A + \overline{\Psi}_\Phi \gamma^\mu \Psi_\Phi \right],$$

and from (3.40)

$$\delta_\varepsilon \Psi_A = i\varepsilon \hat{L}_\Phi \Psi_\Phi, \qquad \delta_\varepsilon \Psi_\Phi = i\varepsilon \hat{L}_A (i\partial_{A_1}) \Psi_A.$$

If (3.54) is not satisfied, we would have to add new terms to action and investigate their s-supersymmetry. Since with every additional constraint we reduce the dimension of configuration space this process terminates. If the remaining dynamical system is non-trivial then is by construction s-supersymmetric. What would remain is to derive Poisson bracket constraint algebra and divide the constraints into constraints of the first and the second class. In our case we have to prove that

$$\int d^4 x \, tr \left[ \delta_\varepsilon K^\mu \tilde{\Phi}_\mu + K^\mu \delta_\varepsilon \tilde{\Phi}_\mu \right] = 0 . \tag{3.55}$$

Using equations of motion and anti-commutativity of $\tilde{\Phi}_\mu$ we obtain that

$$\delta_\varepsilon K^\mu \tilde{\Phi}_\mu = i\varepsilon \left( tr \left[ \overline{\overline{\Psi}}_\Phi \left( \overline{\overline{\hat{L}}}_\Phi \gamma^\mu - \gamma^\mu \hat{L}_A \right) \left( i \vec{\partial}_{A_1} \right) \Psi_A \right] \tilde{\Phi}_\mu + c.c. \right),$$

$$\delta_\varepsilon \tilde{\Phi}_\mu = \delta_\varepsilon \Psi \left( (L_\Phi \Phi)_1 \right) = i\varepsilon \Psi \left( (L_\Phi L_A \nabla_{A_1} \Phi)_1 \right),$$

where derivatives in $\overline{\overline{\hat{L}}}_\Phi$ act on $\overline{\overline{\Psi}}_\Phi$. We leave the details of derivation of (3.55) to another publication. Here we will just mention that chiral multiplet supercurrent current for non-linear realization of s-supersymmetry in the interacting case

$$J_\varepsilon^\mu = tr \left[ \overline{\Psi}_A \left( -\vec{\partial}_{-A_1} \right) \gamma^\mu \hat{L}_\Phi \Psi_\Phi + \overline{\overline{\Psi}}_\Phi \overline{\overline{\hat{L}}}_\Phi \gamma^\mu \left( \vec{\partial}_{A_1} \right) \Psi_A \right], \tag{3.56}$$

is conserved on-shell. Indeed, using equations of motion derived from non-constrained action (3.29) and the fact that on-shell

$$\left( \vec{\partial}_{A_1} \right) \hat{L}_\Phi \Psi_\Phi \equiv \Psi \left( \nabla_{A_1} L_\Phi \Phi_L \right) = 0,$$

we obtain that

$$\partial_\mu J_\varepsilon^\mu = 0 . \tag{3.57}$$

In general s-supersymmetry (3.40) is incompatible with gauge symmetry because of the presence of the non-gauge covariant term $\delta_{A_1} A_R$ in (3.40). However, the situation becomes more favorable if we consider physical gauge degrees of freedom on-shell and replace the linear Lorentz gauge-fixing condition $\delta A_R = 0$ with its gauged Lorentz analog $\delta_{A_1} A_R = 0$. If $i\vec{\partial}_{A_1} \Psi_\Phi = 0$ and $\delta_{A_1} A_R = 0$ then we obtain

$$\delta_\varepsilon A_R = i\sqrt{2} \varepsilon P_+ \Phi_L , \qquad \delta_\varepsilon \Phi_L = -i\sqrt{2} \varepsilon P_L d_{A_1} A_R . \tag{3.58}$$

We see that while for $p \neq 1$ the quantities on the left and the right hand sides of (3.58) transform the same, for example the two sides of

$$(\Phi_L)_{p+1} + \delta_\varepsilon (\Phi_L)_{p+1} = (\Phi_L)_{p+1} - \varepsilon\, P_L\, d_{A_1} (A_R)_p \qquad (3.59)$$

transform in the fundamental representation as required, for $p = 1$ $(\Phi_L)_1$, the 1-form component of $\Phi_L$, and $(\delta_\varepsilon \Phi_L)_1$ transform differently. The same happens for $(A_R)_1$ and its s-supersymmetric variation. We conclude that gauge interactions break non-linear realization of s-supersymmetry as well. Only if $\Phi_L$ transforms in the adjoint representation can we expect compatibility of non-linear s-supersymmetry and gauge symmetry for all $p$. This, in fact, happens for $\Phi_L$ coupled to gravity. Therefore, gravitational interaction considered as a gauge field does not break s-supersymmetry. But that could have been already derived from the fact that ungauged exterior derivatives on a curved space-time manifold already contain covariant interaction with gravitation via coupling to metric contained in the covariant contraction $\delta$ and Hodge $*$-operator.

In order to understand how to incorporate the Lagrangian corresponding to action (3.23) or (3.29) into the quantum gauge-fixed Lagrangian we have to consider expressions for covariant divergence for the factors of the gauge group of the SM. These are given by

$$\delta_{A_1} A_1 = -\partial^\mu A_\mu + ig\, A^\mu A_\mu \qquad \text{for } U(1),$$

$$(\delta_{A_1} A_1)^a = -\partial^\mu A_\mu^a \qquad \text{for } SU(2), \qquad (2.19)$$

$$(\delta_{A_1} A_1)^a = -\partial^\mu A_\mu^a + i\frac{g}{4} d^{abc} A_\mu^b A^{\mu c} \qquad \text{for } SU(N),\ N \geq 3.$$

We see that for $SU(2)$ we can extend linear realization of s-supersymmetry of the free field Lagrangian to non-linear realization of the interacting Lagrangian without modification of the gauge-fixing condition. However, for $U(1)$ and $SU(N), N \geq 3$ the Lorentz gauge-fixing condition $\partial^\mu A_\mu = 0$ has to be modified to

$$(\delta_{A_1} A_1) = 0. \qquad (3.60)$$

For $U(1)$ such modification, in effect, is adding a gauge symmetry breaking mass term. For $SU(N), N \geq 3$ we have a more complicated additional term, which, nevertheless, is still quadratic. It is interesting that such non-linear gauge-fixing conditions with additional quadratic terms appear when one deals with the problem Gribov copies [48, 49].

The last step to adapt the quantum gauge-fixed Lagrangian to s-supersymmetry of interacting action (3.29) is to compute the corresponding Faddeev-Popov determinant. We obtain the gauge-fixed Lagrangian with partial non-linear s-supersymmetry (3.40) of action (3.29)

$$\mathcal{L} = \mathcal{L}_{int} + \mathcal{L}_{gf} + \mathcal{L}_{FP},$$

$$\mathcal{L}_{int} = -\frac{1}{2} tr\left(\alpha\, d_{A_1} A, d_{A_1} A\right) + tr\left(\alpha\, \Phi, \left(d_{A_1} - \delta_{A_1}\right) \Phi\right),$$

$$\mathcal{L}_{gf} = -\frac{1}{\zeta_p} tr\left(\alpha\, \delta_{A_1} A_p, \delta_{A_1} A_p\right), \tag{3.61}$$

$$\mathcal{L}_{FP} = 2 tr\left(\alpha\, \overline{c}, D_{FP} c\right),$$

where the Faddeev-Popov operator $D_{FP}$ is given by

$$D_{FP} = \left(\partial^\mu - 2ig\, A^\mu\right) D_\mu \qquad \text{for } U(1),$$

$$D_{FP}{}^{ab} = \partial^\mu D_\mu{}^{ab} \qquad \text{for } SU(2), \tag{3.62}$$

$$D_{FP}{}^{ac} = \left(\partial^\mu \delta^{ab} - i\frac{g}{2} d^{adb} A^{\mu d}\right) D_\mu{}^{bc} \qquad \text{for } SU(N),\ N \geq 3.$$

As we mentioned above, for $SU(2)$ Lagrangian (3.61) is identical to Lagrangian (3.1).

## 4. $SU(N) \times U(1)$ and $SU(N_3) \times SU(N_2) \times U(1)$ Scalar Supersymmetry

We will now consider non-Abelian gauge groups of practical interest, namely $G = SU(N) \times U(1)$ and $G = SU(N_3) \times SU(N_2) \times U(1)$. For simplicity, in this section we will assume that $N_g = 4$. As we will presently see, the $U(1)$ factor in $G$ is actually required to realize s-supersymmetry. In this section we will set the $U(1)$ generator to one instead of $1/\sqrt{2}$ in previous sections. The difference can be absorbed in the field redefinition and is not essential for discussion.

To introduce s-supersymmetry it is first necessary to equalize the number of the bosonic and the fermionic degrees of freedom. To match the degrees of freedom for non-Abelian gauge groups, in addition to promoting gauge field 1-form $A_1$ to an arbitrary real inhomogeneous real difform $A$, we have to assign $A$ and $\Phi_L$ to appropriate representations of the gauge group. We have to keep in mind that $\Phi_L$ represents fermions and is an arbitrary chiral complex inhomogeneous difform. Further, the physical (anti-)Dirac components $\psi^A$ of bi-spinors $\Psi = \Psi(\Phi_L)$, must transform in the fundamental representation of $SU(N)$ for each non-trivial factor in $G$ for each generation index $A = 1, \ldots, N_g$, while the real gauge difform $(A_R)_1$ must transform in the adjoint representation of the non-trivial factors of $G$ up to a gauge transformation and in the trivial representation of the $U(1)$ factor.

From these requirements we obtain that the simplest choice with equal number of degrees of freedom for real $A$, where we dropped index $R$, and chiral $\Phi_L$ is when $\Phi_L$ transforms in $N \times \overline{N}$ of $SU(N)$, the direct product of fundamental and anti-fundamental representations of the $SU(N)$ factor. This representation is obtained if we use spinbein decomposition of $\Phi_L$ with (anti)-Dirac fields $\psi_L^{aA}$ and spinbein $\eta^{aA}$ given by

$$\Phi_L^{ab} = tr(Z\Psi_L^{ab}), \quad \Psi_L^{ab} = \psi_L^{aA}\overline{\eta}^{bA}, \quad \overline{\eta}^{aA} = \Gamma^{AB}\overline{\eta}^{aB}, \quad \psi_{L,R} = \frac{1}{2}(1 \mp \gamma^5)\psi, \quad (4.1)$$

with $\psi_L^{aA}$, $\eta^{aA}$ transforming in the $N$ of $G$. This $N \times \overline{N}$ representation is not reducible because $\psi_L^{aA}$ and $\eta^{aA}$ are unrelated. However, for real difform $A$, which also has to transform $N \times \overline{N}$ of $G$, $N \times \overline{N}$ is reducible and separates into its irreducible components according to

$$A^{ab} = \frac{1}{N}B\delta^{ab} + W^{ab}, \quad B = tr\,A, \quad tr\,W = 0, \quad a,b = 1,\ldots,N, \quad (4.2)$$

where $B$ transforms in the trivial representation of $U(1)$, while $W^{ab}$ transforms in the $(N^2-1)$-dimensional adjoint representations of $SU(N)$ up to a gauge transformation. Note that since $\eta^a$ are physical objects that are not observable as quantum fields [28, 30] our representation assignment matches the physical degrees of freedom but does not match the observable degrees of freedom. In fact, in our massless example the number of the observable gauge degrees of freedom per helicity state is $N^2$ for bosons, while for fermions it is $N_g \cdot N = 4N$.

We can now write down s-supersymmetry transformations for left chiral scalar supermultiplet with free field action in the $\xi = 2$ gauge (the right chiral case is completely analogous) and action

$$S_0 = \int d^4x\, tr(\alpha\,(d-\delta)A_R, (d-\delta)A_R) + \int d^4x\, tr(\alpha\,\Phi_L, (d-\delta)\Phi_L), \quad (4.3)$$

where in terms of the irreducible bosonic field components $B$, $W$, of difform $A$ the bosonic part of (4.3) becomes

$$S_g^0 = \int d^4x\, tr(\alpha\,(d-\delta)B, (d-\delta)B) + \int d^4x\, tr(\alpha\,(d-\delta)W, (d-\delta)W). \quad (4.4)$$

Following the same steps as for (3.25) we obtain our fourth main result that (4.3) is invariant under

$$\delta A_R = i\varepsilon L_\Phi \Phi_L, \qquad \delta \Phi_L = -i\varepsilon L_A (d-\delta) A_R, \quad (4.5)$$

or, equivalently, under

$$\delta \Phi_L = -i\varepsilon L_A (d-\delta)\left(\frac{1}{N}\delta^{pq} B + W^{pq}\right),$$

$$\delta B = i\varepsilon L_\Phi\, tr\,\Phi_L, \qquad \delta W^{pq} = i\varepsilon L_\Phi\left(\Phi_L{}^{pq} - \frac{1}{N}\delta^{pq} tr\,\Phi_L\right),$$

(4.6)

where $\varepsilon$ is an infinitesimal real Grassmann parameter. As before from Noether's theorem we obtain a single charge $Q$, with $Q^2 = 0$. Note that s-supersymmetry mixes the irreducible components $B$, $W$ of $A$ and the left chiral (anti-)Dirac spinors $\psi_L^{aA}$ together. By taking trace of both sides of (4.6) we obtain a subgroup of the s-supersymmetry group that mixes the bosonic fields $B$ of the $U(1)$ factor of $SU(N)\times U(1)$ and $tr\,\Phi_L$. The non-linear realization of s-supersymmetry for interacting action can be obtained in complete analogy with the $U(1)$ case of the previous section by replacing in (4.3-6) operators $d$, $\delta$ and the conversion operators with their gauged versions. We will not go in detail for lack of space.

It is not difficult to extend our results to $G = SU(N_3)\times SU(N_2)\times U(1)$, which together with already explored cases of $U(1)$ and $SU(N)\times U(1)$, can be used to construct the Lagrangian for s-supersymmetric bi-spinor reformulation of the SM. The expansion of the bosonic difform into irreducible components $B$, $W$, $G$, $tr\,W = 0$, $a,b = 1,\ldots,N_2$, $tr\,G = 0$, $i,k = 1,\ldots,N_3$, becomes

$$A^{(ai)(bk)} = \frac{1}{N_2}\frac{1}{N_3} B\delta^{ab}\delta^{ik} + \frac{1}{N_3} W^{ab}\delta^{ik} + \frac{1}{N_2}\delta^{ab} G^{ik},$$

$$tr_{2\times 3} A = B, \qquad tr_3 A = \frac{1}{N_2} B\delta^{ab} + W^{ab}, \qquad tr_2 A = \frac{1}{N_3} B_1 \delta^{ik} + G_1^{ik}, \qquad (4.7)$$

$$d_{A_1} = d - i\,gB_1 \wedge - i\,g'W_1 \wedge - i g_s G_1 \wedge,$$

where index of the trace indicates over which group factor the trace is taken. Similarly, the fermionic difform $\Phi_L$ and its bi-spinor field $\Psi(\Phi_L)$ also acquire additional representation indices and now are described by

$$\Phi_L^{(ai)(bk)} = tr\left(Z\,\Psi_L^{(ai)(bk)}\right). \qquad (4.8)$$

The interacting BRST Lagrangian for the bi-spinor reformulation of massless SM can be written as

$$\mathcal{L} = \mathcal{L}_g + \mathcal{L}_f + \mathcal{L}_{gh}, \qquad (4.9)$$

$$\mathcal{L}_g = \left(\alpha(d-\delta)B, (d-\delta)B\right) - tr_2\left(\alpha(d_{W_1}-\delta)W, (d_{W_1}-\delta)W\right)$$

$$- tr_3\left(\alpha(d_{G_1}-\delta)G, (d_{G_1}-\delta)G\right), \qquad (4.10)$$

$$\mathcal{L}_f = tr\left(\alpha\, \Phi_R^l, \left(d_1^{\,l} - \delta_1^{\,l}\right)\Phi_R^l\right) + tr\left(\alpha\, \Phi_L^l, \left(d_{1\times 2}^{\,l} - \delta_{1\times 2}^{\,l}\right)\Phi_L^l\right) +$$
$$+ tr\left(\alpha\, \Phi_R^q, \left(d_{1\times 3}^{\,q} - \delta_{1\times 3}^{\,q}\right)\Phi_R^q\right) + tr\left(\alpha\, \Phi_L^q, \left(d_{1\times 2\times 3}^{\,q} - \delta_{1\times 2\times 3}^{\,q}\right)\Phi_L^q\right),$$
(4.11)

where $\mathcal{L}_g$ is the gauge sector Lagrangian, $\mathcal{L}_f$ is the fermionic sector Lagrangian, and $\mathcal{L}_{gh}$ contains gauge-fixing and ghost terms, which we will not write out for brevity. The right (left) lepton covariant derivative $d_1^{\,l}$ ($d_{1\times 2}^{\,l}$) and the right (left) quark covariant derivative $d_{1\times 3}^{\,q}$ ($d_{1\times 2\times 3}^{\,q}$) are given by

$$d_1^{\,l} = d - i\, gB_1 \wedge ,$$
$$d_{1\times 2}^{\,l} = d - i\, gB_1 \wedge -i\, g'W_1 \wedge ,$$
$$d_{1\times 3}^{\,q} = d - i\, gB_1 \wedge -i\, g_s G_1 \wedge ,$$
$$d_{1\times 2\times 3}^{\,q} = d - i\, gB_1 \wedge -i\, g'W_1 \wedge -i\, g_s G_1 \wedge .$$
(4.12)

Following the same steps as for (4.5, 4.6) we obtain various s-supersymmetries involving the bosonic difforms and the left- and the right-handed fermionic difforms. For example, (4.9) is invariant under two sets of independent s-supersymmetry transformations involving subgroups $SU(N_2)\times U(1)$ and $SU(N_3)\times U(1)$ of $SU(N_3)\times SU(N_2)\times U(1)$. These are obtained by reducing the bosonic difforms to transform in the representations of the two subgroups. The corresponding reduction of the fermionic difform $\Phi_L^{(ai)(bk)}$ is achieved by taking appropriate traces. The two s-supersymmetries use

$$\Phi_L^2 = tr_3 \Phi_L, \qquad \Phi_L^{2\,ab} = \Phi_L^{(ak)(bk)},$$
$$\Phi_L^3 = tr_2 \Phi_L, \qquad \Phi_L^{3\,ik} = \Phi_L^{(ai)(ak)},$$
(4.13)

where the first or second line in (4.13) corresponds to the first or second line in (4.11). We obtain explicitly that the two free field s-supersymmetries for left-handed fermions are given by

$$\delta B = i\varepsilon\, L_\Phi\, tr\, \Phi_L^2, \qquad \delta W^{ab} = i\varepsilon\, L_\Phi \left(\Phi_L^{2\,ab} - \frac{1}{N_2}\delta^{ab} tr_2\, \Phi_L^2\right),$$
$$\delta \Phi_L^{2\,(ai,bk)} = -i\varepsilon\, L_A (d-\delta)\delta^{ik}\left(\frac{1}{N}\delta^{ab} B + W^{ab}\right),$$
(4.14)

and

$$\delta B = i\varepsilon L_\Phi\, tr\, \Phi_L^3, \qquad \delta G^{ik} = i\varepsilon L_\Phi \left( \Phi_L^{3\,ik} - \frac{1}{N_2}\delta^{ab} tr_3\, \Phi_L^3 \right),$$

$$\delta \Phi_L^{3\,(ai,bk)} = -i\varepsilon L_A (d-\delta)\delta^{ab}\left( \frac{1}{N}\delta^{ik} B + G^{ik} \right), \tag{4.15}$$

where $\varepsilon$ is an infinitesimal real Grassmann parameter. As before, from Noether's theorem we obtain a single charge for each transformation. Similarly, one can obtain various s-supersymmetries involving bosonic difforms and fermionic difforms obtained from $\Phi_{L,R}^l$ and $\Phi_{L,R}^q$ by taking appropriate traces.

The generalization of the results of this section to the interacting case is analogous to $U(1)$ case. Here we will only outline the full derivation. For simplicity, we will consider $SU(N) \times U(1)$ example only.
Instead of (4.3 – 6) in the $\xi = 2$ gauge we obtain

$$S = \int d^4 x\, tr\big(\alpha\, \nabla_{A_1} A_R, \nabla_{A_1} A_R\big) + \int d^4 x\, tr\big(\alpha\, \Phi_L, \nabla_{A_1} \Phi_L\big),$$

$$S_b = \int d^4 x\, tr\big(\alpha\, \nabla_{B_1} B, \nabla_{B_1} B\big) + \int d^4 x\, tr\big(\alpha\, \nabla_{W_1} W, \nabla_{W_1} W\big),$$

$$\delta A_R = i\varepsilon L_\Phi\, \Phi_L, \qquad \delta \Phi_L = i\varepsilon L_A \nabla_{A_1} A_R,$$

$$\delta \Phi_L = i\varepsilon L_A \nabla_{A_1} \left( \frac{1}{N}\delta^{pq} B + W^{pq} \right),$$

$$\delta B = i\varepsilon L_\Phi\, tr\, \Phi_L, \qquad \delta W^{pq} = i\varepsilon L_\Phi \left( \Phi_L^{pq} - \frac{1}{N}\delta^{pq} tr\, \Phi_L \right).$$

If we now apply non-linear s-supersymmetry transformation to the interacting action then, as in the $U(1)$ case, the first part of variation of the action $\delta_\varepsilon^{free} S$ where gauge difform in $\nabla_{A_1}$ is not varied vanishes similarly to the free field case. For the residual part we obtain again

$$\delta_\varepsilon^{res} S = \delta_\varepsilon S_b + \delta_\varepsilon S_f, \tag{4.16}$$

where the bosonic and fermionic components are given by

$$\delta_\varepsilon S_b = \frac{g}{2}\varepsilon \int d^4 x \left( tr\left[ \overline{\Psi}_A \overline{\Psi}((L_\Phi \Phi_L)_1) \vec{\partial}_{A_1} \Psi_A - \overline{\Psi}_A \vec{\partial}_{-A_1} \Psi((L_\Phi \Phi_L)_1) \Psi_A \right] \right), \tag{4.17}$$

$$\delta_\varepsilon S_f = -i\frac{g}{2}\varepsilon \int d^4 x \left( tr\left[ \overline{\overline{\Psi}}_\Phi \left( \overline{\Psi}((L_\Phi \Phi_L)_1) + \Psi((L_\Phi \Phi_L)_1) \right) \Psi_\Phi \right] \right), \tag{4.18}$$

where now trace involves Lie algebra indices as well, and $\Psi_A \equiv \Psi^a(A_R) T^a$, $\Psi_\Phi \equiv \Psi^a(\Phi_L) T^a$ with generators $T^a$ decomposed into irreducible $U(1)$ and $SU(N)$ parts according to

$$T^a = \{1, \tau^a\}, \quad T^a \hat{\Phi}^a_\mu = \frac{1}{\sqrt{2}} \Phi_\mu + \tau^a \Phi^a_\mu, \qquad \mu = 0, \cdots, 3. \tag{4.19}$$

Correspondingly, we can force vanishing of the residual variation by imposing the conditions that are essentially the same as (3.48)

$$tr\left[\left(\vec{\partial}_{A_1} \Psi^a{}_A \overline{\overline{\Psi}}^c{}_A + \Psi^a{}_A \overline{\overline{\Psi}}^c{}_A \overleftarrow{\overline{\partial}}_{A_1}\right) \gamma^\mu\right] tr\left[T^a T^b T^c\right] \hat{\Phi}^b_\mu = 0,$$

$$tr\left[\overline{\overline{\Psi}}^a{}_\Phi \gamma^\mu \Psi^c{}_\Phi\right] tr\left[T^a T^b T^c\right] \hat{\Phi}^b_\mu = 0, \tag{4.20}$$

where the first trace in (4.20) is taken over the Lorentz indices and the second over the Lie algebra indices. Obviously, the conditions (4.20) are satisfied when we impose in addition to (3.48) for the irreducible $U(1)$ component of $A_R$ an obvious generalization of conditions (3.48) on $\Psi^a{}_{A,\Phi}$, the $SU(N)$ irreducible components of expansion of $\Psi_{A,\Phi}$ in the Lie algebra basis. We will consider these constraints in more detail elsewhere.

## 5. Scalar Superstring

In the last section, we will show briefly that exact s-supersymmetry has a realization as a global supersymmetry of a string action. The action is a supersymmetric version of the bi-spinor string action described in [50]. It gives an example of exact s-supersymmetry that is symmetry of the whole action, not only of its part. Representation of fermions in string theories via bi-spinor difforms was first described in [50] and then rediscovered in [51], were also the supersymmetry of the bi-spinor string action was investigated. The emphasis in [51] was to find a bi-spinor string action that has conformal invariance of the standard bosonic string. This can be done at the expense of adding new field variables and the end result is a somewhat complicated modification of the string action whose relevance to string theory was unclear.

We will keep the simple form of supersymmetric string action with bi-spinors and instead of modifying it to ensure conformal invariance off-shell, we will consider it as a part of the more general quantum action of the bi-spinor gauge theory that we described above that uses constraints that ensure conformal invariance on-shell. String supersymmetry with a somewhat different Lagrangian $\propto tr\left(\left(d B^A, d B_A\right) + \left(F^A, dF_A\right)\right)$ with real $F^A$ was considered in [52, 53]. However, such Lagrangians cannot describe chiral spinors. Therefore, the significance of such Lagrangians is not clear.

In more detail, consider a collection of complex commuting and anticommuting 2-forms $B^A$ and $F^A$, $A = 0, \ldots, D-1$, transforming in some representation of a gauge group and defined on a two dimensional manifold with metric $g_{\mu\nu}$, $\mu, \nu = 0, 1$ that is imbedded into $D$-dimensional Minkowski space-time $M_D$ with metric $\eta_{AB}$. Assume that $B^A$ and $F^A$ transform in the same representation of a gauge group. Then the action

$$S = \int \sqrt{-g}\, d^2x\, \eta_{AB}\, tr\left((\alpha(d-\delta)B^A, (d-\delta)B^B) + (\alpha F^A, (d-\delta)F^B)\right) \quad (5.1)$$

is globally both gauge invariant and supersymmetric under the transformation

$$\delta B^A = \varepsilon F^A, \quad (5.2)$$

$$\delta F^A = -\varepsilon^*(d-\delta)B^A, \quad (5.3)$$

where trace is over the gauge group representation indices. Expanding

$$B^A = B_0^A + B_\mu^A dx^\mu + (1/2)B_2^A \varepsilon_{\mu\nu} dx^\mu \wedge dx^\nu, \quad (5.4)$$

and taking into account that

$$dB_0^A = \partial_\mu B_0^A dx^\mu, \quad \delta B_0^A = 0, \quad (\alpha\, dx^\mu, dx^\nu) = -g^{\mu\nu}, \quad (5.5)$$

we find that (5.1) contains two bosonic strings described by $\mathrm{Re}\, B_0^A$, $\mathrm{Im}\, B_0^A$ coupled to D pairs of Dirac-anti-Dirac spinors. The fields $B_2^A$ are non-dynamical.

In the alternative, one can use left or right chiral difforms for fermions and real difforms for bosons. Then only one bosonic string described by real $B_0^A$ is contained in (5.1). It is coupled to chiral fermionic fields.

Because of the presence of $\delta$ in action (5.1) it is not conformally invariant. However, if we impose the standard Lorentz constraint on the bosonic difform

$$\delta B^A = 0, \quad (5.6)$$

then conformal invariance of the bosonic part of string action (5.1) is restored. We then obtain the constrained action in the form similar to that considered in [52, 53]

$$S = \int \sqrt{-g}\, d^2x\, \eta_{AB}\, tr\left((\alpha\, dB^A, d\, B^B) + (\alpha F^A, dF^B) + (\alpha\, dF^A, F^B)\right). \quad (5.7)$$

Constraint (5.6) does not affect the scalar 0-form $B_0^A$ that describes the standard bosonic string. At the same time it eliminates some of the longitudinal modes of $B_1^A$, which is a reasonable physical requirement for string theory that should describe massless excitations.

One possibly interesting direction of further research for bi-spinor string action is to consider the gauged version of action (5.1) written for a general p-brane. We obtain the gauged p-brane bi-spinor string action

$$S = \int \sqrt{-g}\, d^p x\, \eta_{AB}\, tr\!\left(\!\left(\alpha\!\left(d_{B_1}-\delta_{B_1}\right)\!B^A,\left(d_{B_1}-\delta_{B_1}\right)\!B^B\right)+\left(\alpha F^A,\left(d_{B_1}-\delta_{B_1}\right)\!F^B\right)\!\right). \tag{5.8}$$

This action, when coupled with non-linear Lorentz constraint $\delta_{B_1} B^A = 0$, is $U(1)$ gauge invariant and is s-supersymmetric under the non-linear realization of s-supersymmetry along the lines of chiral s-supersymmetry (3.40). It has an interesting dual interpretation. On the one hand, we can consider it as a string action for a string propagating in $D$-dimensional Minkowski space-time with metric $\eta_{AB}$. On the other hand we can consider it as a bi-spinor gauge theory on curved background with global non-compact internal symmetry described by Lorentz transformations. Thus four-dimensional bi-spinor gauge theory on curved background with global $D$-dimensional Lorentz symmetry may be considered equivalent to a 4-brane propagating in the $D$-dimensional Minkowski space-time.

## 6. Summary

In summary, we presented a new realization of supersymmetry acting in the space of commuting and anticommuting difforms. We proved it for free filed case and outline the proof for the interacting case. Implementation of scalar supersymmetry relieves supersymmetric models from requiring that each observed particle must have a new superpartner particle. In its simplest version interacting massless s-supersymmetry leads to the existence of a Higgs-like field and thus scalar supersymmetric modifications of the SM can be done with the observed particle spectrum of the SM.

S-supersymmetry, in the form discussed here, is at the same time a more general notion then the standard supersymmetry and a more restrictive notion. It is more general, because it can be defined on any smooth space-time manifold. It is more restrictive, because it does not admit extended supersymmetry. Further, it can only be defined if fermionic matter is represented by bi-spinors, instead of Dirac spinors.

Interestingly, s-supersymmetry requires the appearance of $U(1)$ factor in the gauge group. It suggests why the left- and the right-handed fermions couple differently to $SU(2)$ gauge fields: coupling to right-handed fields requires another set of gauge fields. These have not been observed. Whether s-supersymmetry leads to gauge coupling unification is an open question. However, it is clear that at least some of the benefits of supersymmetry of the free action should be inherited in the interacting theory in the ultraviolet behavior of Feynman loop integrals.

Both in the free in interacting cases s-supersymmetry is broken by gauge interactions. This breakdown is beneficial, because in both cases s-supersymmetry is restored in the high energy limit for asymptotically free theories. Thus even the simplest realizations of s-supersymmetry have two most salient features of softly broken standard supersymmetry built in from the beginning. It is broken at low energies allowing for differences in masses of the particles, while at high energies it is restored allowing for the benefits of supersymmetry be applied to divergent graphs.

Finally, we showed that exact global s-supersymmetry can be realized in a superstring action. Since s-supersymmetry is not equivalent to the standard supersymmetry, s-supersymmetric string might provide an alternative setting for construction of the theory of quantum gravity interacting with gauge fields and bi-spinor fermionic matter. How bi-spinor s-supersymmetric action fits into the standard superstring classification and what is its critical dimension are open questions.

It is the Dirac spinors rather then bi-spinors that are the mathematical objects used in the Standard Model to describe fermions. We showed that the use of bi-spinors instead of Dirac spinors brings certain advantages. It allows one to avoid the use of torsion when describing coupling of fermions to gravity, provides a realization of supersymmetry that is more compact then the standard one, fixes the number of generations to three, and leads to essentially unique texture of lepto-quark mixing. All these might be an indication that bi-spinors offer a more fitting description of quantum fermionic matter.

**Appendix A: Conventions, Differential Geometry, Z-basis, and Spinbeins**

In the Appendix we will work with a (pseudo)-Euclidean manifold $M$ of dimension $n$ endowed with metric $g = \{g_{\mu\nu}\}$ with signature $s$, which is the number of the negative eigenvalues of $g$. To reduce the general results to phenomenologically interesting four-dimensional Minkowski space-time $M_4$ one needs to substitute in the formulas $n = 4$, $g_{\mu\nu} = diag(1,-1,-1,-1)$, $s = 3$.

We will use the following index conventions: Capital Latin letters $A, B, \ldots$ are reserved for the fermion generations, capital Latin letters $I, J, \ldots$ denote tangent space indices, lower case Latin letters $a, b, \ldots$ are for adjoint Lie algebra representation, lower case Latin letters $i, j, \ldots$ for its fundamental representation, lower case Greek letters $\alpha, \beta, \ldots$ for spinor indices, while $\mu, \nu, \ldots$ for Lorentz tensor indices. In our expressions $tr$ denotes trace over the representation indices and, where necessary, contains additional trace over the spinor indices.

The basic notions of differential geometry that we need are the standard operations with difforms on a manifold [54, 55], a special basis in the space of difforms, called the $Z$-basis, which is used to define bi-spinors [13], and the spinbein decomposition of bi-spinors [28], which is used to extract Dirac spinors from bi-spinors.

Given $M$ with coordinates $x^\mu$, a difform $A$ in the coordinate basis (c-basis) is defined as a sum of homogeneous difforms of degree $p$ with values in some vector space, for example, in $\mathfrak{g}$ the Lie algebra or its representation of the gauge group $G$ where product of two elements of is defined as their commutator

$$A(x) = \sum_{p=0}^{n} A_p(x),$$

$$A_p(x) = A_{|\mu_1 \cdots \mu_p|}(x) dx^{|\mu_1} \wedge \cdots \wedge dx^{\mu_p|} = \frac{1}{p!} A_{\mu_1 \cdots \mu_p}(x) dx^{\mu_1} \wedge \cdots \wedge dx^{\mu_p},$$

(A.1)

where $\wedge$ is the exterior product of differentials and $|\mu_1 \cdots \mu_p|$ is a permutation of indices $\mu_1 \cdots \mu_p$ with increasing order. In bi-spinor formalism such difforms play the role of the fields of the standard (quantum) field theory. Equivalently one can define a difform in the basis formed by exterior products of coframe 1-forms $e^I = e_\mu^I dx^\mu$

$$A_p(e) = A_{|I_1 \cdots I_p|}(x)\, e^{|I_1|} \wedge \cdots \wedge e^{|I_p|}, \qquad g(e^I, e^J) = \eta^{IJ}.$$

Additional basic differential-geometric constructs that we need are the main automorphism $\alpha$, the main antiautomorphism $\beta$, and contraction $(.,.)$ of a $p$-form $A_p$ with a $q$-form $B_q$ that from a pair $\{A_p, B_q\}$ generates a $|p-q|$-form $C_{|p-q|}$. These are given by

$$\alpha A_p = (-1)^p A_p, \qquad\qquad \beta A_p = (-1)^{p(p-1)/2} A_p, \qquad (A.2)$$

$$(A_p, B_q)_{\mu_1 \cdots \mu_{q-p}} = \frac{1}{p!} \left(A^{\mu_1 \cdots \mu_p}\right)^\dagger B_{\mu_1 \cdots \mu_q}, \qquad p \leq q,$$

$$(A.3)$$

$$(A_p, B_q)_{\mu_1 \cdots \mu_{p-q}} = \frac{1}{q!} \left(A_{\mu_1 \cdots \mu_p}\right)^\dagger B^{\mu_1 \cdots \mu_q}, \qquad p \geq q,$$

where $A^+$ is either complex or Hermitean conjugate of the coefficients of $A$, and space-time indices of difforms are raised with $g^{\mu\nu} = (g^{-1})^{\mu\nu}$, which on $M_4$ becomes $g^{\mu\nu} = diag(1,-1,-1,-1)$. When $p = 1 \leq q$ the definition (A.3) coincides with the definition of inner product of $B_q$ with the vector field dual to $A_1$. If necessary a trace in (A.3) is taken over the indices of Lie algebra representation. In particular for

$$A_p(x) = dx^{|\mu_1|} \wedge \cdots \wedge dx^{|\mu_p|}, \qquad B_q(x) = dx^{|\nu_1|} \wedge \cdots \wedge dx^{|\nu_q|},$$

we have

$$(A_p, B_q) = g^{\mu_1 \nu_1} \cdots g^{\mu_p \nu_p} dx^{|\nu_{p+1}|} \wedge \cdots \wedge dx^{|\nu_q|}, \qquad p < q,$$

$$(A_p, B_q) = g^{\mu_1 \nu_1} \cdots g^{\mu_p \nu_p}, \qquad p = q,$$

$$(A_p, B_q) = g^{\mu_1 \nu_1} \cdots g^{\mu_q \nu_q} dx^{|\mu_{q+1}|} \wedge \cdots \wedge dx^{|\mu_p|}, \qquad p > q.$$

The exterior derivative $d$ is defined by

$$d : A_p \to A_{p+1}, \quad dA_n = 0, \qquad dA_p = \partial_\nu A_{\mu_1 \cdots \mu_p} dx^\nu \wedge dx^{\mu_1} \wedge \cdots \wedge dx^{\mu_p}. \quad (A.4)$$

Commutativity of derivatives in (A.4) leads to $d^2 = 0$.

The space Hodge star operator $*$, $* : A_p \to A_{n-p}$, is defined by

$$1 \wedge *1 = \varepsilon_n, \qquad \varepsilon_n = e^1 \wedge \cdots \wedge e^n, \qquad (A.5)$$

where $\varepsilon_n$ is the volume element. Its action on cotangent space coefficients of a difform is given by

$$*: A_p \to A_{n-p}, \quad \left((*A)_{n-p}\right)_{I_1...I_{n-p}} = \frac{1}{p!(n-p)!} \varepsilon_{I_1...I_{n-p}}{}^{J_1...J_p} A_{J_1...J_p}, \tag{A.6}$$

$$\varepsilon^{0...n-1} = 1, \quad \varepsilon_{I_1...I_n} = -\varepsilon^{I_1...I_n}.$$

Note that our normalization of $\varepsilon^{0...n-1} = 1$ is the standard in quantum field theory, while in theories of gravity often $\varepsilon^{0...n-1} = -1$ is used. The square of Hodge star is proportional to unity

$$** = (-1)^{p(n-p)+s}. \tag{A.7}$$

For $M_4$ we have

$$** = (-1)^{p+1} = -\alpha, \tag{A.8}$$

$$\varepsilon^{0123} = 1, \quad \varepsilon_{\mu_1...\mu_4} = -\varepsilon^{\mu_1...\mu_4}. \tag{A.9}$$

Note that definition of Hodge $*$-operator involves metric but exterior derivative does not use it. From $d$ and $*$ the covariant divergence operator, or coderivative, $\delta$ is defined by

$$\delta: A_p \to A_{p-1}, \quad \delta A_0 = 0, \tag{A.10}$$

$$\delta = (-1)^{n(p+1)+s+1} * d *, \tag{A.11}$$

which for $M_4$ reduces to $\delta = *d*$. Operator inherits dependence on metric from Hodge $*$. From $d^2 = 0$ and (A.8, A.11) we obtain $\delta^2 = 0$. Very useful for us will be operator $\star$, which we will call the chiral star operator, defined by

$$\star = -i * \alpha \beta. \tag{A.12}$$

Although it does not belong to the standard set of definitions of differential geometry, it is very useful for our purposes; we will use it to define chirality of difforms. On $M_4$ chiral star is self-adjoint with respect to (A.14) and satisfies

$$\star\star = 1, \qquad \star^+ = \star. \tag{A.13}$$

Applied to $A_2$, difforms of the second degree, the chiral star operator reduces to the standard definition of chiral 2-forms in terms of selfdual or anti-selfdual 2-forms. Note that our definition of chiral star operator differs from the extension of the Euclidean chiral star operator [13] extended to Minkowski space-time by the presence of $\alpha$. Its presence ensures that $(d-\delta)P_{L,R} = P_{R,L}(d-\delta)$, which corresponds to $\partial \gamma^5 = -\gamma^5 \partial$ in the Z-basis. For operator defined in [13] one obtains $(d-\delta)P_{L,R} = P_{L,R}(d-\delta)$, which is not the property that is needed to establish correspondence between coordinate-free and Z-basis description of bi-spinors.

We define a scalar product $\langle A, B \rangle$ of difforms $A, B$ on manifolds without boundary by linearity from

$$\langle A_p, B_q \rangle = \int tr\left[(\alpha A_p^+) \wedge * B_q\right] = \delta_{pq} \int \sqrt{g}\, dx^0 \wedge \cdots \wedge dx^{n-1} (\alpha A_p, B_q). \quad (A.14)$$

This scalar product is non-zero if $|p-q|=1$ and space-time manifold has a boundary. Note that because of the presence of automorphism $\alpha$ it is $-\delta$ that is the adjoint of $d$ with respect to scalar product (A.14) and, therefore, $(d-\delta)$ is self-adjoint. For Euclidean space-time definition (A.14) must be modified by removing automorphism $\alpha$. The presence of $\alpha$ in (A.14) is motivated by the reduction of difforms to bi-spinors and extraction from them of algebraic Dirac spinors in such a way that the standard Dirac action is obtained. We will describe the decomposition and the extraction next.

We will now introduce a special basis, the Z-basis, in the space of difforms [13] and establish the connection between difforms, antisymmetric tensors, and bi-spinors. Given a manifold $M$ one can define a set of Dirac $2^{[n/2]}$-dimensional $\gamma$-matrices,

$$\gamma^\mu = \{\gamma^\mu_{\alpha\beta}\}, \quad \{\gamma^\mu, \gamma^\nu\} = 2g^{\mu\nu}. \quad (A.15)$$

Their cotangent space analogs are given by

$$\gamma^I = \{\gamma^I_{\alpha\beta}\}, \quad \gamma^I = e^I_\mu \gamma^\mu, \quad \{\gamma^I, \gamma^J\} = 2\eta^{IJ}, \quad (A.16)$$

where $e^I_\mu$ are coefficients of frame one-forms $e^I$, $e^I = e^I_\mu dx^\mu$ and $\eta^{IJ} = \eta_{IJ} = diag(1,-1,\cdots,-1)$. For 2k-dimensional $M$ we can define n-dimensional analog of four-dimensional $\gamma^5$-matrix, which we denote by $\gamma^5$, by $\gamma^5 = i\gamma^0 \cdots \gamma^{n-1}$. We will be working in the chiral basis for $\gamma$-matrices on $M_4$ given by

$$\gamma^\mu = \begin{pmatrix} 0 & \sigma^\mu \\ \bar{\sigma}_\mu & 0 \end{pmatrix}, \quad \gamma^5 = \begin{pmatrix} -1 & 0 \\ 0 & 1 \end{pmatrix}, \quad \gamma^0 = \begin{pmatrix} 0 & 1 \\ 1 & 0 \end{pmatrix},$$

$$\sigma^\mu = (1, \sigma^k), \quad \bar{\sigma}^\mu = (1, -\sigma^k),$$

$$\sigma^1 = \begin{pmatrix} 0 & 1 \\ 1 & 0 \end{pmatrix}, \quad \sigma^2 = \begin{pmatrix} 0 & -i \\ i & 0 \end{pmatrix}, \quad \sigma^3 = \begin{pmatrix} 1 & 0 \\ 0 & -1 \end{pmatrix},$$

$$\sigma^i \sigma^j = i\varepsilon_{ijk}\sigma^k.$$

The defining property of the $Z$-basis, $Z = \{Z_{\alpha\beta}\}$, is that operator $(d-\delta)$, which in the mathematical literature is called the signature operator [57, 58], takes the form of the Dirac operator

$$(d-\delta)Z = Z(i\gamma^\mu \partial_\mu). \tag{A.17}$$

On $M_4$ the $Z$-basis is a $4\times 4$ matrix of difforms. Any difform $A$ can be represented in the $Z$-basis as

$$A = tr(Z\,\Psi(A)), \qquad A^+ = tr((\beta Z)\overline{\overline{\Psi}}(A)), \tag{A.18}$$

where $\Psi(A) = \{\Psi_{\alpha\beta}(A)\}$ are the coefficients of the representation and the trace is over the $\gamma$-matrix indices. Using (A.15-16) we obtain an explicit expression for $Z$ and its (A.14) adjoint

$$Z = \sum_p \gamma_{\mu_p}\cdots\gamma_{\mu_1} dx^{|\mu_1} \wedge\cdots\wedge dx^{\mu_p|}, \tag{A.19}$$

$$\overline{C}Z \equiv Z^+ = \beta\gamma^0 Z \gamma^0. \tag{A.20}$$

Since difforms do not depend on the basis in which they are defined, the coefficients $A_{\mu_1\cdots\mu_p}(A)$ of $A$ in the c-basis and the coefficients $\Psi_{\alpha\beta}(A)$ of $A$ in the $Z$-basis represent the same mathematical object. Also the transformation properties of the two sets of coefficients can be derived from basis independence of $A$: under Lorentz transformation $x \to \Lambda x$ the set $\{A_{\mu_1\cdots\mu_p}(A)\}$ transforms as a collection of covariant antisymmetric tensors, while $\Psi_{\alpha\beta}(A)$ transforms as

$$\Psi(A) \to S(\Lambda)\,\Psi(A)\,S(\Lambda)^{-1}, \tag{A.21}$$

where $S(\Lambda)$ is the spinor representation of the appropriate local Lorentz group. Transformation (A.21) by definition is the transformation law for bi-spinors, for they transform as a product of a Dirac spinor and its Dirac conjugate. Thus, we can identify the space of all $\Psi$ with the space of bi-spinors. We described the construction of the local $Z$-basis. The basis can be globalized using the standard procedure. Since we can always define difforms on a smooth manifold and the associated set of cotangent space $\gamma$-matrices, bi-spinors can be defined on any smooth manifold. The same does not apply to Dirac spinors. Dirac spinors can be defined only if there exists a spin structure, which is not guaranteed for an arbitrary smooth $M$ [41].

On $M_4$ relations between the two sets of coefficients $\{A_{\mu_1\cdots\mu_p}(A)\}$ and $\Psi_{\alpha\beta}(A)$ are derived using (A.18) and the completeness relations for $\gamma$-matrices given by

$$tr\left(\left[\gamma^{|\mu_1} \cdots \gamma^{\mu_p|}\right]\left[\gamma^{|\nu_1} \cdots \gamma^{\nu_q|}\right]^+\right) = 2^{n/2}\, \delta^{pq}\, \delta^{\mu_1\nu_1} \cdots \delta^{\mu_q\nu_p}, \tag{A.22}$$

$$\sum_p \left[\gamma^{|\mu_1} \cdots \gamma^{\mu_p|}\right]^*_{\alpha\beta}\left[\gamma^{|\mu_1} \cdots \gamma^{\mu_p|}\right]_{\gamma\delta} = 2^{n/2}\, \delta_{\alpha\gamma}\, \delta_{\beta\delta}. \tag{A.23}$$

The first relation is easy to prove by inspection, the second follows from the fact that (A.22) implies that $2^n \times 2^n$ matrix $\left(\gamma^{|\mu_1} \cdots \gamma^{\mu_p|}\right)_{(\alpha\beta)}$ with multi-indexes $|\mu_1,\ldots,\mu_p|$ and $(\alpha,\beta)$ is unitary. Using (A.22-23) we obtain

$$\left(A_p\right)_{|\mu_1\cdots\mu_p|}(A) = tr\left(\left(\gamma_{\langle\mu_p} \cdots \gamma_{\mu_1\rangle}\right)\Psi(A)\right), \tag{A.24}$$

$$\Psi(A_p) = 2^{-n/2} \sum_p \gamma^{|\mu_1} \cdots \gamma^{\mu_p|} A_{|\mu_1\cdots\mu_p|}, \tag{A.25}$$

where $\langle\mu_p,\ldots,\mu_1\rangle$ denotes permutation of indices with decreasing order.

One property of $Z = \{Z_p\}$ that is needed to define chirality of difforms on even-dimensional $M$ is

$$-i * \alpha\beta Z_p = Z_{n-p}\, \gamma^5, \qquad \gamma^5 = i\gamma^0 \cdots \gamma^{n-1}. \tag{A.26}$$

Using the property we obtain for any difform $A = tr(Z\Psi)$ on $M$

$$-i * \alpha\beta A = tr\left(Z\, \gamma^5 \Psi(A)\right). \tag{A.27}$$

We can now define chiral difforms $A_{L,R}$ on $M$ using projection operators $P_{L,R}$ constructed with the use of chiral star operator (A.12)

$$A_{L,R} = P_{L,R}A, \qquad P_{L,R} = \frac{1}{2}(1 \mp \star), \qquad P_{L,R}^2 = 1,\ P_L P_R = 0. \tag{A.28}$$

Note that on $M$ chiral projection operators (A.28) can be defined only if $A$ is complex-valued. This can be seen from definition of chiral star operator in (A.12). The situation is different for Euclidean manifolds, where $(*\beta)^2 = 1$ and one can define real chiral difforms [13]. From (A.26-28) we obtain that in the $Z$-basis the coefficients of chiral difforms are chiral bi-spinors

$$\left(1 \pm \gamma^5\right)\Psi(A_{L,R}) = 0. \tag{A.29}$$

For convenience, a complete compact list of operations and commutators of the operators that appear in bi-spinor gauge theories on four-dimensional $M$ is given in Appendix B and Tables 1 – 4. The relations listed there also hold if we replace $d, \delta$ with their gauged versions $d_{A_1}, \delta_{A_1}$.

The last ingredient we need to describe s-supersymmetry is the spinbein decomposition of bi-spinors that extracts algebraic Dirac spinors from bi-spinor $\Psi$ transforming in some representation of the gauge group: $\Psi = \{\Psi^{ab}\}$. For $N_g = 4$ generations the extraction is done by using a spinbein $\eta^{aA}$ $a = 1, \ldots, N_\eta$, $A = 1, \ldots, 4$, that is a multiplet of four commuting normalized Dirac spinors transforming in some $N_\eta$-dimensional representation of the gauge group $G$

$$\bar{\bar{\eta}}_\alpha^{aA} \eta_\alpha^{aB} = \delta^{AB}, \tag{A.30}$$

$$\bar{\bar{\eta}}^{aA} = \Gamma^{AB} \bar{\eta}^{aB}, \qquad \Gamma^{AB} = diag(1,\ 1, -1, -1), \tag{A.31}$$

where $\bar{\eta}$ denotes the Dirac conjugate of $\eta$. It follows from (A.30) that in the basis for $\gamma$- matrices where $\gamma^0 = \Gamma$ all spinbeins for an Abelian gauge group are given by elements of group $U(2,2)$. Spinbein decomposition of a bi-spinor is defined as the anzatz [28]

$$\Psi^{ab} = \psi^{aA} \bar{\bar{\eta}}^{Ab}, \tag{A.32}$$

where four generations of Dirac spinors $\psi^{aA}, a = 1, \ldots, N_\psi$, transform in a $N_\psi$-dimensional representation of $G$, which is not necessarily the same as that for the spinbein. From (A.29) we see that definite chirality of bi-spinor $\Psi^{ab}$ is inherited by its Dirac spinor constituents $\psi^{aA}$. However, the form of spinbein decomposition (A.30-32) also implies that there are no right chiral bi-spinors: equation $\Psi(1 \pm \gamma^5) = 0$ has no solutions.

The number of generations in (A.32) can be reduced from four to three or less if one uses a set of generally covariant constraints $\det \Psi^{ab} = 0$ for each pair $a, b$. The second known method to reduce the number of generations contained in a bi-spinor is the decomposition of $\Psi$ into minimal ideals of the associated Clifford algebra [12]. However, while coinciding with ours on $M_4$, this method is not generally covariant and cannot be used in the presence of gravity. Explicit reduction can be achieved by the use of degenerate spinbein with modified normalization condition. For example, reduction of free action from $N_g = 4$ to $N_g = 3$ generations follows from assuming

$$\bar{\bar{\eta}}_\alpha^{aA} \eta_\alpha^{aB} = diag(1, 1, 1, 0). \tag{A.33}$$

Such degenerate spinbeins for any number $N < 4$ of generations can be obtained from spinbeins satisfying (A.30) by zeroing one or more of the eigenvalues of non-degenerate spinbeins by appropriate matrix factor.

In order to set normalization of the action for gauge fields we choose parameter $c$ such that

$$c\int tr[\alpha F \wedge *F] = -\frac{1}{4}\int d^n x\, F_{\mu\nu}{}^a F^{\mu\nu a}\,. \tag{A.34}$$

Since $F = \frac{1}{2} F_{\mu\nu}{}^a \tau^a dx^\mu \wedge dx^\nu$ we obtain

$$c\int tr[\alpha F \wedge *F] = \frac{c}{8}\int F_{\mu\nu}{}^a F_{\rho\sigma}{}^a dx^\mu \wedge dx^\nu \wedge *(dx^\rho \wedge dx^\sigma)\,. \tag{A.35}$$

Using (A.6) we obtain

$$*(dx^\rho \wedge dx^\sigma) = \frac{1}{4}\varepsilon^{\rho\sigma}{}_{\lambda\eta} dx^\lambda \wedge dx^\eta\,,$$

and

$$c\int tr[\alpha F \wedge *F] = \frac{c}{32}\int F_{\mu\nu}{}^a F_{\rho\sigma}{}^a \varepsilon^{\rho\sigma}{}_{\lambda\eta}\varepsilon^{\mu\nu\lambda\eta} d^4 x = -\frac{c}{4}\int F_{\mu\nu}{}^a F^{\mu\nu a} d^4 x\,. \tag{A.36}$$

where we used $\varepsilon^{\rho\sigma}{}_{\lambda\eta}\varepsilon^{\mu\nu\lambda\eta} = -2(g^{\rho\mu}g^{\sigma\nu} - g^{\rho\nu}g^{\sigma\mu})$. We finally obtain that $c = 1$.

Given two arbitrary difforms $F, H$, in the $Z$-basis we can write scalar product (A.14) as

$$\langle F_p, H_q\rangle = \delta_{pq}\int \sqrt{g}\, d^n x\, tr\left[\overline{\overline{\Psi}}(F_p)\Psi(H_q)\right],\quad \overline{\overline{\Psi}}(F_p) = \gamma^0\, \Psi^+(F_p)\,\gamma^0\,, \tag{A.37}$$

where we used $*\alpha\beta Z_p = -Z_{n-p}\Gamma^5$ from (A.26), hence $*\alpha\beta Z = \beta * Z = -Z\Gamma^5$, $\Gamma^5 \equiv \gamma^0 \cdots \gamma^{n-1} = -i\gamma^5$,

$$(\alpha Z^+{}_{\alpha\beta} \wedge *Z_{\gamma\delta})_n = (\alpha Z_{\alpha\beta}, Z_{\gamma\delta})\, dx^0 \wedge \cdots \wedge dx^{n-1} = (\gamma^0)_{\gamma\alpha}(\gamma^0)_{\beta\delta}\, dx^0 \wedge \cdots \wedge dx^{n-1}\,. \tag{A.38}$$

Note that adjointness with respect to (A.37) is not equivalent to adjointness of a matrix. For example, $\gamma^5$ is a Hermitean matrix. As an operator acting on bi-spinors with scalar product (A.37) it is actually anti-Hermitean. We will always specify when adjointness is defined with respect to (A.14) or, equivalently, with respect to (A.37).

The appearance of $\gamma^0$ in (A.37) is the result of the presence of automorphism $\alpha$ in the definition of the scalar product (A.14). It is not present in the Euclidean case. After spinbein anzatz (A.32) we obtain an equivalent representation of the scalar product in terms of Dirac spinor components

$$\langle F, H\rangle = \int \sqrt{g}\, d^n x\, tr\left[\overline{\overline{\psi}}^A(F)\psi^A(H)\right],\qquad \overline{\overline{\psi}}^A(F) = \Gamma^{AB}\overline{\psi}^B(F)\,. \tag{A.39}$$

where $\overline{\overline{\Psi}}(F)$ and $\overline{\overline{\psi}}^A(F)$ are the bi-spinor conjugates of a bi-spinor and Dirac spinor, respectively, while $\overline{\psi}^B$ denotes the standard Dirac conjugation.

To set our normalization conventions for Lie algebras, consider a semisimple compact gauge group $G$ and its Lie algebra $\mathfrak{g}$ with orthonormally chosen basis generators $\tau^a$, $a = 1,\ldots, M$ of some representation of $\mathfrak{g}$ such that

$$[\tau^a, \tau^b] = i f^{abc} \tau^c, \quad tr(\tau^a \tau^b) = \frac{1}{2} \delta^{ab}, \tag{A.40}$$

where $f^{abc}$ are the structure constants of $g$. For $G = U(1)$ $M = 1$, while for $G = SU(N)$ we have $M = (N^2 - 1)$. For $SU(N)$ a unique completely symmetric invariant $d^{abc}$ can be defined via anti-commutator of matrices of the orthonormal basis (A.40) of the fundamental representation

$$\{t^a_F, t^b_F\} = \frac{1}{N} \delta^{ab} + d^{abc} t^c_F, \quad d^{abc} = 0 \quad \text{for } N = 2. \tag{A.41}$$

The invariant $d^{abc}$ is of phenomenological interest, because it defines the anomaly coefficient $A(r)$ of a representation $\{t^a_r\}$ by

$$tr[t^a_r \{t^b_r, t^c_r\}] = \frac{1}{2} A(r) d^{abc}, \quad A(r_F) = 1. \tag{A.42}$$

To ensure the uniformity in the treatment of the Abelian and non-Abelian cases we will set the trivial generator for the $U(1)$ gauge group to be

$$\tau^1 = 1/\sqrt{2}, \quad tr(\tau^1)^2 = \frac{1}{2}. \tag{A.43}$$

This means that to obtain physical $U(1)$ gauge fields in our formalism one needs to rescale the $U(1)$ coupling by a factor $\sqrt{2}$:

$$g = \frac{1}{\sqrt{2}} (g)_{phys}. \tag{A.44}$$

**Appendix B: Commutation Relations of Basic Operators**

|  | $\overline{C}$ | $\alpha$ | $\beta$ | $*$ | $d$ | $\delta$ |
|---|---|---|---|---|---|---|
| $\overline{C}$ | 1 | $\alpha\overline{C}$ | $\beta\overline{C}$ | $*\overline{C}$ | $d\overline{C}$ | $\delta\overline{C}$ |
| $\alpha$ | $\overline{C}\alpha$ | 1 | $\beta\alpha$ | $*\alpha$ | $-d\alpha$ | $-\delta\alpha$ |
| $\beta$ | $\overline{C}\beta$ | $\alpha\beta$ | 1 | $*\alpha\beta$ | $d\alpha\beta$ | $-\delta\alpha\beta$ |
| $*$ | $\overline{C}*$ | $\alpha*$ | $\alpha\beta*$ | $-\alpha$ | $-\delta\alpha*$ | $d\alpha*$ |
| $d$ | $\overline{C}d$ | $-\alpha d$ | $-\alpha\beta d$ | $-*\alpha\delta$ | 0 | $-\delta d - \Delta$ |
| $\delta$ | $\overline{C}\delta$ | $-\alpha\delta$ | $\alpha\beta\delta$ | $*\alpha d$ | $-d\delta - \Delta$ | 0 |

Table 1. Commutators of the primary operators for (pseudo)-Euclidean space-time dimension $n = 4$ and signature $s = 3$.

|  | $\nabla_-$ | $\star$ | $P_L$ | $P_R$ | $P_+$ | $P_-$ |
|---|---|---|---|---|---|---|
| $\overline{C}$ | $\nabla_-\overline{C}$ | $-\star\overline{C}$ | $P_R\overline{C}$ | $P_L\overline{C}$ | $P_+$ | $P_-$ |
| $\alpha$ | $-\nabla_-\alpha$ | $\star\alpha$ | $P_L\alpha$ | $P_R\alpha$ | $P_+\alpha$ | $P_-\alpha$ |
| $\beta$ | $\nabla_+\alpha\beta$ | $-\star\beta$ | $\frac{1}{2}(1-\alpha\star)\beta$ | $\frac{1}{2}(1+\alpha\star)\beta$ | $P_+\beta$ | $P_-\beta$ |
| $*$ | $\overline{C}*$ | $\star\alpha*$ | $\frac{1}{2}(1-\alpha\star)*$ | $\frac{1}{2}(1+\alpha\star)*$ | $P_+*$ | $P_-*$ |
| $d$ | $\overline{C}d$ | $\star\delta$ | $\frac{1}{2}(d-\star\delta)$ | $\frac{1}{2}(d+\star\delta)$ | $P_+d$ | $P_-d$ |
| $\delta$ | $\overline{C}\delta$ | $\star d$ | $\frac{1}{2}(\delta-\star d)$ | $\frac{1}{2}(\delta+\star d)$ | $P_+\delta$ | $P_-\delta$ |

Table 2. Commutators of the primary and the secondary operators for (pseudo)-Euclidean space-time dimension $n = 4$ and signature $s = 3$.

|  | $\overline{C}$ | $\alpha$ | $\beta$ | $*$ | $d$ | $\delta$ |
|---|---|---|---|---|---|---|
| $\nabla_-$ | $\overline{C}\nabla_-$ | $-\alpha\nabla_-$ | $-\alpha\beta\nabla_+$ | $-*\beta\nabla_+$ | $-\delta d$ | $d\delta$ |
| $\star$ | $-\overline{C}\star$ | $\alpha\star$ | $-\beta\star$ | $*\alpha\star$ | $\delta\star$ | $d\star$ |
| $P_L$ | $\overline{C}P_R$ | $\alpha P_L$ | $\frac{1}{2}\beta(1-\alpha\star)$ | $\frac{1}{2}*(1-\alpha\star)$ | $\frac{1}{2}(d-\delta\star)$ | $\frac{1}{2}(\delta-d\star)$ |
| $P_R$ | $\overline{C}P_L$ | $\alpha P_R$ | $\frac{1}{2}\beta(1+\alpha\star)$ | $\frac{1}{2}*(1+\alpha\star)$ | $\frac{1}{2}(d+\delta\star)$ | $\frac{1}{2}(\delta+d\star)$ |
| $P_+$ | $P_+$ | $\alpha P_+$ | $\beta P_+$ | $*P_+$ | $dP_+$ | $\delta P_+$ |
| $P_-$ | $-P_-$ | $\alpha P_-$ | $\beta P_-$ | $*P_-$ | $dP_-$ | $\delta P_-$ |

Table 3. Commutators of the secondary and the primary operators for (pseudo)-Euclidean space-time dimension $n=4$ and signature $s=3$.

|  | $\nabla_-$ | $\star$ | $P_L$ | $P_R$ | $P_+$ | $P_-$ |
|---|---|---|---|---|---|---|
| $\nabla_-$ | $\Delta$ | $-\star\nabla_-$ | $P_R\nabla_-$ | $P_L\nabla_-$ | $P_+\nabla_-$ | $P_-\nabla_-$ |
| $\star$ | $-\nabla_-\star$ | $1$ | $-P_L$ | $P_R$ | $P_+\star$ | $-P_-\star$ |
| $P_L$ | $\nabla_-P_R$ | $-P_L$ | $P_L$ | $0$ | $\frac{1}{2}(P_+-iP_-\star)$ | $\frac{1}{2}(P_-+iP_+\star)$ |
| $P_R$ | $\nabla_-P_L$ | $P_R$ | $0$ | $P_R$ | $\frac{1}{2}(P_++iP_-\star)$ | $\frac{1}{2}(P_--iP_+\star)$ |
| $P_+$ | $\nabla_-P_+$ | $\star P_+$ | $\frac{1}{2}(P_+-i\star P_-)$ | $\frac{1}{2}(P_++i\star P_-)$ | $P_+$ | $0$ |
| $P_-$ | $\nabla_-P_-$ | $-\star P_-$ | $\frac{1}{2}(P_-+i\star P_+)$ | $\frac{1}{2}(P_--i\star P_+)$ | $0$ | $P_-$ |

Table 4. Commutators of the secondary operators for (pseudo)-Euclidean space-time dimension $n=4$ and signature $s=3$.